\documentclass[aps,prl,amsmath,amssymb,superscriptaddress,twocolumn,10pt]{revtex4-2}
\usepackage[english]{babel}
\usepackage[utf8]{inputenc}
\usepackage{amsfonts}
\usepackage[T1]{fontenc}
\usepackage[pdftex]{graphicx}
\usepackage{fancyhdr}
\usepackage{hyperref}
\hypersetup{
    bookmarks=false,         
    unicode=false,          
    pdftoolbar=false,        
    pdfmenubar=true,        
    pdffitwindow=false,     
    pdfstartview={FitH},
    pdfnewwindow=true,      
    colorlinks=true,       
    linkcolor=black,          
    citecolor=blue,        
    filecolor=magenta,      
    urlcolor=blue           
}

\usepackage{url}
\usepackage[titletoc,title]{appendix}
\usepackage{epstopdf}
\usepackage{amsmath}
\usepackage{amssymb}
\usepackage{dsfont}
\usepackage{mathtools}
\usepackage{bbold}
\usepackage{xcolor}
\usepackage{tikz}
\usetikzlibrary{arrows.meta}

\graphicspath{{./Figures/}}

\newcommand{\tr}{{\rm tr}}

\newcommand{\ket}[1]{|#1\rangle}

\newcommand{\change}[1]{#1}

\newcommand{\mytitle}{Anomalous transport in the kinetically constrained quantum East-West model}

\begin{document}

\title{\mytitle}
\author{Pietro Brighi} 
\affiliation{Faculty of Physics, University of Vienna, Boltzmanngasse 5, 1090 Vienna, Austria}
\author{Marko Ljubotina} 
\affiliation{Institute of Science and Technology Austria (ISTA), Am Campus 1, 3400 Klosterneuburg, Austria}
\date{\today}

\begin{abstract}
    We study a chaotic particle-conserving kinetically constrained model, with a single parameter which allows us to break reflection symmetry.
    Through extensive numerical simulations we find that the domain wall state shows a variety of dynamical behaviors from localization all the way to ballistic transport, depending on the value of the reflection breaking parameter.
    Surprisingly, such anomalous behavior is not mirrored in infinite-temperature dynamics, which appear to scale diffusively, in line with expectations for generic interacting models.
    However, studying the particle density gradient, we show that the lack of reflection symmetry affects infinite-temperature dynamics, resulting in an asymmetric dynamical structure factor.
    This is in disagreement with normal diffusion and suggests that the model may also exhibit anomalous dynamics at infinite temperature in the thermodynamic limit. 
    Finally, we observe low-entangled eigenstates in the spectrum of the model, a telltale sign of quantum many body scars. 
\end{abstract}

\maketitle

\textit{Introduction -- }
Out-of-equilibrium properties of many-body systems present one of the central problems in quantum statistical mechanics. 
Of particular interest are the different universality classes of dynamics found in various models. 
Typically, generic chaotic models are expected to behave diffusively~\cite{Prosen_2012, Karrasch_2014, Lux2014, Znidaric2016, Gu_2017, Blake_2017, Friedman_2020, De_Nardis_2020, bertini2021finite, Zeiher2022}, although slower dynamics were observed in disordered systems~\cite{Basko06, Demler2015, Reichman15, Imbrie16, Znidaric2016, Bar_Lev_2017, quasiperiodic2018}. 

Recently however, it was shown that certain chaotic kinetically constrained models (KCMs) can exhibit superdiffusive dynamics at infinite temperature~\cite{Ljubotina2023}. 
Such non-diffusive behavior was observed both in particle~\cite{Vasseur2021} and energy transport~\cite{Ljubotina2023}.
Additionally, anomalous dynamics can also arise at the level of pure states~\cite{Garrahan2015,Garrahan2018,Michailidis2018,Pancotti2020,Marino2022,Brighi2023a}, as in the celebrated PXP model, where certain states show long-lived oscillations in the density of domain walls~\cite{Lukin2017,Michailidis2018}.

Besides anomalous dynamical features, kinetically constrained models, first introduced in the context of classical glasses~\cite{Andersen1984,Andersen1993,Ritort2003}, also host other remarkable phenomena.
These range from Hilbert space fragmentation~\cite{Pollmann2020a,Nandkishore2020,Iadecola2020,Pozsgay2021,Sen2021,Zadnik2021,Motrunic2022} to quantum many-body scars~\cite{Michailidis2018,Turner2018,Bernevig2018,Choi2019,Iadecola2019,Lukin2019,Papic2020,Knolle2020,Knolle2021,Tamura2022,Regnault2022}, defining the novel paradigm of weak ergodicity breaking~\cite{Serbyn2021}.

A paradigamtic example of KCMs is the celebrated quantum East model~\cite{Garrahan2015}.
The quantum East model hosts a localization transition in the ground state~\cite{Pancotti2020,Garrahan2023} and extremely slow dynamics~\cite{Garrahan2015,Garrahan2023}, while its Floquet version has shown localized behavior~\cite{Prosen2023} as well as an exactly solvable point in parameter space~\cite{Garrahan2023a}.
However, the model only has a single conserved charge, the energy, which itself is not present in the Floquet version. 

A recent work~\cite{Brighi2023a} introduced a particle conserving version of the quantum East model.
The combination of $U(1)$ symmetry and kinetic constraints leads to classical and quantum Hilbert space fragmentation, i.e.~fragmentation in an entangled basis~\cite{Motrunic2022}, and to a dramatic effect on dynamics, which show superdiffusive behavior in certain initial states~\cite{Brighi2023a}.

\begin{figure}[b]
    \centering
    \includegraphics[width=.95\columnwidth]{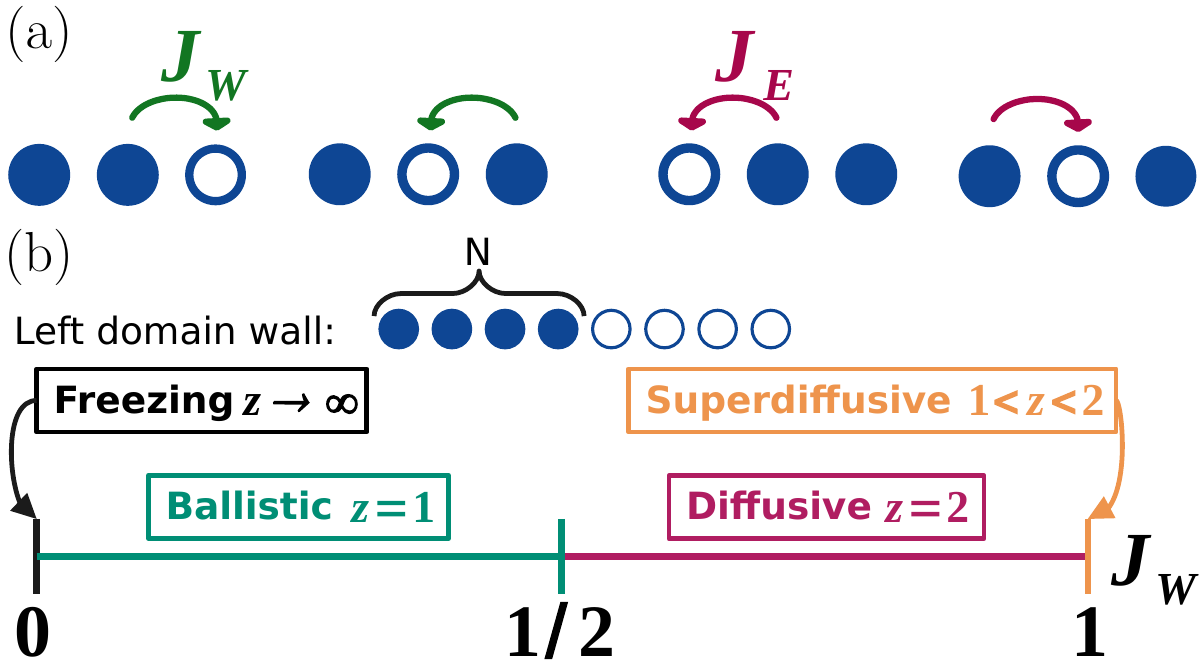}
    \caption{
    \label{Fig:cartoon}
    (a) Particles move with different hopping amplitudes \change{$J_E=1-J_W$} depending on whether the constraint is applied from the left or the right neighbor.
    (b) As \change{$J_W$} is changed from $0$ to $1$, the dynamical behavior of the \textit{left domain wall} state, shown here for the small case of $N=4$ particles and $L=8$ sites, changes dramatically.
    Its dynamics show a transition from fully localized at \change{$J_W=0$} to ballistic for \change{$J_W\leq1/2$} and to the expected diffusive behavior as \change{$J_W$} is increased further.
    However, at the opposite extreme (\change{$J_W=1$}) the dynamics appear to enter an anomalous superdiffusive regime. 
    \change{Here the dynamical exponent $z$ corresponds to that from Eq. \eqref{Eq:dN}, computed only from the LDW state. }
    }
\end{figure}

In this work we explore the interplay between reflection symmetry and kinetic constraints. 
Specifically, we focus on the dynamics of a constrained hopping model, inspired by the particle conserving quantum East model~\cite{Brighi2023a}.
In the original model hopping is allowed between two neighboring sites if the site immediately to the \change{right} of the pair is occupied.
Here we add the reflection-symmetric \textit{West} constrained hopping term, with a potentially different amplitude.
This also allows particle hopping when the nearest neighbor on the \change{left} is occupied, regardless of the state of the \change{right} neighbor.
The addition of this term breaks both classical and quantum fragmentation, allowing us to study infinite-temperature transport in the dominant fragment of the Hilbert space and control the reflection symmetry. 

In spite of the absence of fragmentation and chaotic level spacing statistics, the spectrum of the Hamiltonian still presents intriguing characteristics~\footnote{\label{footnoteSM}See supplementary material for weakly entangled eigenstates, anomalous zero modes, convergence analysis and additional information about the skewness fitting procedure.}.
In particular, we find a large number of zero modes which depends on particle parity and a set of weakly entangled eigenstates reminiscent of quantum many-body scars~\cite{Michailidis2018,Turner2018}.

Studying the dynamics of the domain wall state~\cite{Antal_1999,Gobert_2005}, a simple and experimentally accessible initial state, we discover the presence of a rich phase diagram dependent on the ratio of the two hopping terms.
In our simulations we observe a full spectrum of different dynamical behaviors \change{of the domain wall state} ranging from completely localized to a surprising ballistic behavior, in spite of the chaotic nature of the model.
This suggests that ballistic transport, typically observed in integrable models, which has been related to superconductivity~\cite{Jiang_2019,Dong_2022}, can also be observed in \change{the dynamics of pure states of} certain generic chaotic models. 
Upon increasing the \change{West} hopping term beyond the symmetric point, where the two hopping amplitudes are equal, the state recovers the expected diffusive spreading.
However, when the \change{East} constraint is completely suppressed, we again observe anomalous dynamics, this time with superdiffusive scaling up to the available times.

Using tensor-network methods, we further probe infinite-temperature dynamics which indicate diffusive scaling.
However, analyzing the dynamical structure factor at infinite temperature~\cite{Prosen2019}, we observe long-lived asymmetry.
This finite asymmetry is at odds with normal diffusion and shows another intriguing anomaly of transport in this model.

\begin{figure*} 
    \centering
    \includegraphics[width=.99\textwidth]{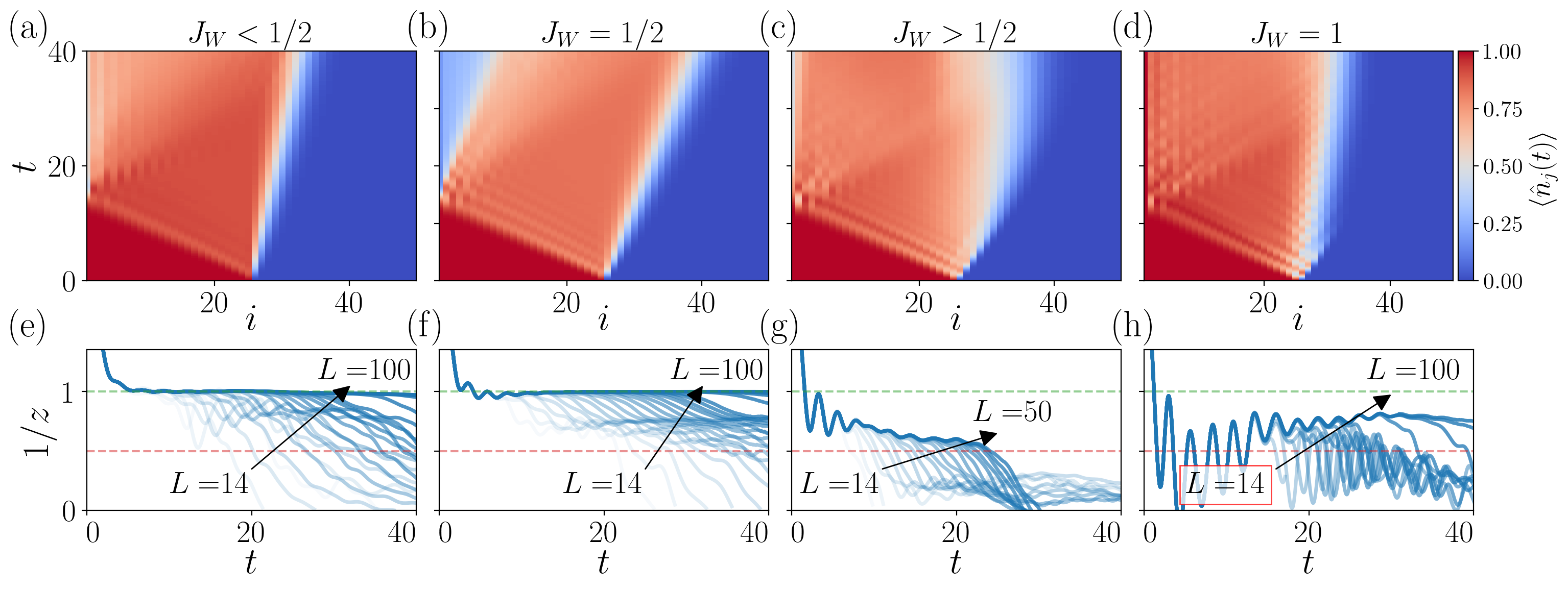}
    \caption{
    \label{Fig:states dynamics} 
    Density dynamics of $\ket{\text{LDW}}$ dramatically changes as \change{$J_W$} is tuned across the symmetric point \change{$J_W=1/2$} (a)-(d).
    While the model is fully chaotic and its infinite-temperature dynamics appear to scale diffusively, for $\change{J_W}\leq1/2$ $\ket{\text{LDW}}$ shows ballistic behavior (a),(b).
    However, as $\change{J_W}>1/2$ (c),(d), the $\ket{\text{LDW}}$ state deviates from the ballistic regime.
    This clearly emerges looking at the instantaneous dynamical exponent $z(t)$, whose inverse is shown in panels (e)-(h).
    For $\change{J_W}\leq1/2$ (e),(f) the dynamical exponent presents a long plateau at $z^{-1}=1$ corresponding to ballistic behavior.
    \change{This eventually changes when finite size effects are observed at a critical time $\tau\propto L$. 
    Therefore we extract $z$ from times where the largest two system sizes have not yet diverged in order to avoid the finite size effects.}
    For $1/2<\change{J_W}<1$ (g), the inverse dynamical exponent quickly decays below $1$, relaxing at long times to a value close to diffusion $z^{-1}=1/2$.
    Finally, as the \change{East} contribution vanishes, $\change{J_W}=1$(h), $\ket{\text{LDW}}$ acquires an unexpected superdiffusive behavior, with the dynamical exponent rapidly oscillating at first, and later approaching a value well above diffusion.
    \change{The dashed lines in the bottom row correspond to diffusive and ballistic behavior (red and green respectively).
    }
    The data were generated for $\change{J_W} = 1/3,1/2,2/3,1$ using exact methods for $L\le24$ and TEBD with bond dimension $\chi=[1536,2048]$ for $L\ge26$.
    }
\end{figure*}

\textit{Model -- }%
We study kinetically constrained hard-core bosons on a one-dimensional lattice of $L$ sites.
Hopping among different sites is allowed only in certain configurations, as encoded in the system Hamiltonian
\begin{equation}
    \label{Eq:H}
    \begin{split}
    \hat{H} &= \change{J_W}\sum_{i=1}^{L-2} \hat{n}_i \left( \hat{c}^\dagger_{i+2}\hat{c}_{i+1}+\text{H.c.}\right) \\
    &+\change{J_E}\sum_{i=1}^{L-2} \left( \hat{c}^\dagger_{i+1}\hat{c}_i+\text{H.c.}\right)\hat{n}_{i+2} = J_E\hat{H}_E+J_W\hat{H}_W,
    \end{split}
\end{equation}
where $\hat{H}_E$($\hat{H}_W$) is the $U(1)$-conserving East(West) Hamiltonian, $\hat{c}^\dagger_{i}$ is the hard-core boson creation operator and $\hat{n}_i = \hat{c}^\dagger_i\hat{c}_i$ is the particle number operator.
The action of the two kinetic constraints is sketched in Figure~\ref{Fig:cartoon}(a).
Particles can hop only if their nearest neighbor \textit{to the left} (\change{West}) or \textit{to the right} (\change{East}) is occupied, with amplitudes \change{$J_W$} and \change{$J_E$} respectively.
Throughout this work, we fix the two hopping parameters such that $J_E + J_W = 1$.
As we will show in the following, varying the parameter \change{$J_W$} leads to the dramatic change in dynamics of the domain wall state~\cite{Antal_1999,Gobert_2005} depicted in Figure~\ref{Fig:cartoon}(b).

Besides being particle-conserving, the Hamiltonian~(\ref{Eq:H}), with periodic boundary conditions, is also translation invariant.
At the \textit{symmetric point} $J_E = J_W = 1/2$ the system is further reflection symmetric.
As opposed to the closely related quantum East model, no additional symmetry emerges and the system does not exhibit Hilbert space fragmentation away from $\change{J_W}\in\{0,1\}$, at least within the half-filling sector on which we focus in this work.
Despite the absence of Hilbert space fragmentation, the analysis of the spectrum and of the eigenstates of the Hamiltonian yields interesting observations~\cite{Note1}.
On one hand the study of the level spacing distribution confirms that the system is overall chaotic.
On the other, we notice the presence of a small number of weakly entangled eigenstates, reminiscent of quantum many-body scars.
Finally, the spectrum presents an anomalously large number of zero modes, which appear only for \textit{even} particle numbers $N=L/2$.

\textit{Domain wall dynamics -- }%
We now focus on the dynamics after a quantum quench in our system with open boundary conditions.
Our protocol consists of initializing the system in the left domain wall state
\begin{equation}
    \label{Eq:psi0}
    \ket{\text{LDW}} =  |\underbrace{\bullet\bullet\bullet\dots\bullet}_{N}\underbrace{\circ\circ\circ\dots\circ}_{L-N}\rangle \quad \begin{array}{l}\hat{n}_i\ket{\bullet} = \ket{\bullet}\\
    \hat{n}_i\ket{\circ} = 0\end{array},
\end{equation}
which at \change{$J_W=0$} is an exact zero energy eigenstate of the Hamiltonian.
We then suddenly switch the \change{West} hopping amplitude to its final value $\change{J_W}\in(0,1]$.

To study the dynamics of $\ket{\text{LDW}}$ we perform numerical simulations over an extensive number of system sizes $L\in [14,100]$, using exact techniques for $L\leq 24$ and approximate matrix-product-state time-evolution using the time-evolving-block-decimation (TEBD) algorithm~\cite{Vidal2003} for $L>24$~\cite{Note1}.
Our analysis focuses on the instantaneous dynamical exponent $z(t)$ defining the dynamical behavior of the state~\change{\footnote{\change{Note that the LDW state is an eigenstate of $\hat{n}_i(0)$, thus computing the expectation value of $\hat{n}_j(t)$ gives us direct access to the correlation function $\hat{n}_i(0)\hat{n}_j(t)$ evaluated in the LDW state. See supplementary material for further details}}}.
For interacting chaotic systems as the one we study, particle spreading \change{in generic high temperature ensembles} is expected to be diffusive $(z = 2)$.
Deviations from this behavior are known in integrable models~\cite{Sutherland2004}, which can present ballistic $(z=1)$ and superdiffusive $(1<z<2)$ transport~\cite{Ilievski2021,Zeiher2022,Znidaric2011,Ljubotina2017,Prosen2018,Prosen2019,bertini2021finite}, and in disordered systems with sub-diffusive dynamics $(z>2)$~\cite{Znidaric2016,Demler2015}.
\change{Here, instead, we focus on a single pure state, similarly to previous studies of the domain wall state in the $XXZ$ chain~\cite{Antal_1999,Gobert_2005}, with weight over the entire spectrum.}
To numerically obtain the instantaneous dynamical exponent, we take the logarithmic derivative of the particle flow from the domain wall 
\begin{align}
    \label{Eq:dN}
    \delta N(t) = \sum_{i=1}^N \langle\hat{n}_i(t=&0)\rangle - \langle\hat{n}_i(t)\rangle, \;\;\delta N(t)\propto t^{1/z} \\
    \frac{1}{z} &= \frac{d\log\delta N}{d\log t}.
\end{align}

In Figure~\ref{Fig:states dynamics} we show the particle dynamics of the domain wall quench (a)-(d) as well as the instantaneous dynamical exponent (e)-(h) at different values of \change{$J_W$}, highlighting the variety of different behaviors in our model.
When $\change{J_W}<1/2$ the dynamics of $\ket{\text{LDW}}$ initially are, as expected, relatively slow due to the combination of dominant \change{East} constraint and large particle density in the left half (a).
Surprisingly, however, the transport exponent shows clear ballistic scaling $z=1$(e).
\change{This behavior persists up to times proportional to the system size due to finite size effects.}
A similar behavior, although with much faster particle spreading, is observed for $\change{J_W}=1/2$, shown in panels (b) and (f).
In both cases the results indicate ballistic dynamics \change{of $\ket{\text{LDW}}$} in the thermodynamic limit. 

As the hopping amplitude is increased even further, entering the regime where \change{$J_W>J_E$}, the domain wall recovers the expected diffusive behavior, as shown by the dynamical exponent $1/z\to1/2$ in Figure~\ref{Fig:states dynamics}(g).
However, at the extreme point \change{$J_W=1$}, where only the \change{West} Hamiltonian participates in the dynamics, the domain wall state acquires yet another unexpected dynamical exponent.
As shown in Figure~\ref{Fig:states dynamics}(h), after a series of initial oscillations damping with systems size, $1/z$ approaches a superdiffusive value, which at the timescales attainable by our simulations is approximately $1/z\approx0.8$. 

While superdiffusion in the particle-conserving East model was recently observed~\cite{Brighi2023a} with exact diagonalisation, here we discover that introducing the additional West constraint and tuning the asymmetry between the two yields the rich dynamical \textit{phase diagram} for the domain wall state shown in Figure~\ref{Fig:cartoon}(b).
In particular, the presence of a ballistic state in an otherwise chaotic model is highly atypical~\cite{Znidaric2019}. 
Finally, we mention that for \change{$J_W\ll J_E$} we notice a striking difference in dynamics depending on the particle number $N$ (akin to the number of zero modes discussed previously).
For even $N$, particles are confined within a small region and cannot explore the full lattice, while for odd $N$ they spread ballistically through the whole chain~\cite{Note1}.

\textit{Persistent asymmetry at infinite temperature -- }%
To further characterize the dynamics in the model, we analyze the dynamical exponent of mixed states close to infinite temperature $\rho_0 = \otimes_i \rho_0^{(i)}$,
\begin{equation}
    \label{Eq:rho0}
    \begin{split}
    \rho_0^{(i)} &= \begin{pmatrix}
        1/2 + \mu(i) & 0 \\
        0 & 1/2 - \mu(i)
    \end{pmatrix}, \;\; \mu(i) = \begin{cases}
        \mu_0 \;\; i\leq L/2 \\
        -\mu_0 \;\; i>L/2
    \end{cases}
    \end{split}
\end{equation}
with $\mu_0\ll 1$~\cite{Ljubotina2017}.
Using TEBD, we simulate the dynamics of a system of $L=512$ sites in a wide parameter range $\change{J_W}\in[0.15,0.5]$ and $\mu\in[0.001,0.1]$.

\begin{figure}[b]
    \centering 
    \includegraphics[width = .95\columnwidth]{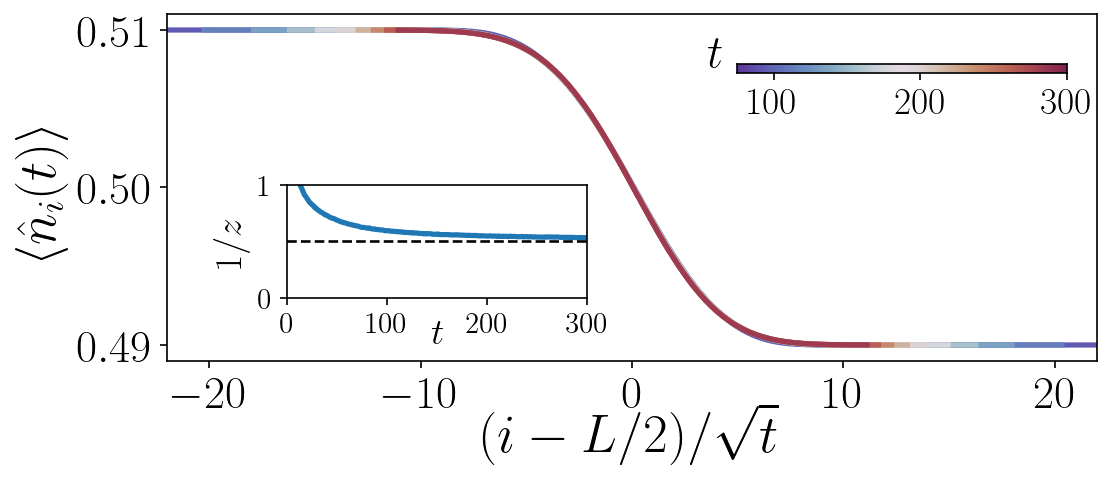}
    \caption{Mixed state density profiles at different times $t\in [75,300]$ collapse on one another upon rescaling the x-axis by $\sqrt{t}$ suggesting diffusive scaling at infinite temperature.
    The diffusive scaling is further confirmed by the dynamical exponent approaching $z^{-1}\approx0.5$ at late times, as shown in the inset.
    The data shown here are for $\change{J_W} = 1/2,\mu = 0.01$ and $\chi = 384$. 
    \label{Fig:diffusion}
    }
\end{figure}

First, we focus on the dynamical exponent $z$.
To get an estimate of its value, we perform a collapse of the density profiles at different times, shown in Figure~\ref{Fig:diffusion}.
Given the dynamical exponent, density dynamics are expected to be captured by a scale-invariant function $f(x/t^{1/z})$.
The different curves in Figure~\ref{Fig:diffusion} collapse on one another within a broad range of times as the space-axis is rescaled by $\sqrt{t}$, suggesting diffusive scaling.
This observation is further confirmed by the dynamical exponent approaching $z=2$ at late times, as reported in the inset.
Similar results are obtained for other values of the hopping parameters~\cite{Note1}.
The usual diffusive behavior, then, seems to be recovered at high temperature.
We note that tiny but non-vanishing faster-than-diffusive corrections would be difficult to identify and thus cannot be ruled out.
Therefore, there could be other states that share the same ballistic behavior as the $\ket{\text{LDW}}$ state, so long as they remain a sufficiently small or vanishing fraction of the Hilbert space size. 

In a diffusive system, particle spreading is expected to be symmetric around the central chemical potential step $\mu$.
Our model, however, is inherently \textit{asymmetric} and could deviate from this behavior.
To characterize this possible asymmetry, we analyze the particle density gradient $\Delta n_{i,i+1} = |\langle \hat{n}_i\rangle_\mu - \langle \hat{n}_{i+1}\rangle_\mu|$, which is related to the dynamical structure factor $S(x,t)=\langle \hat{n}_{x}(t)\hat{n}_0(0)\rangle=\lim_{\mu\to0}\frac 1 \mu \Delta n_{x,x+1}$~\cite{Prosen2019}.
Here $\langle\hat{O}\rangle_\mu$ represents the expectation value of the operator with the weak domain wall initial state from Eq.~\eqref{Eq:rho0}. 

\begin{figure}[t!]
    \centering
    \includegraphics[width=.95\columnwidth]{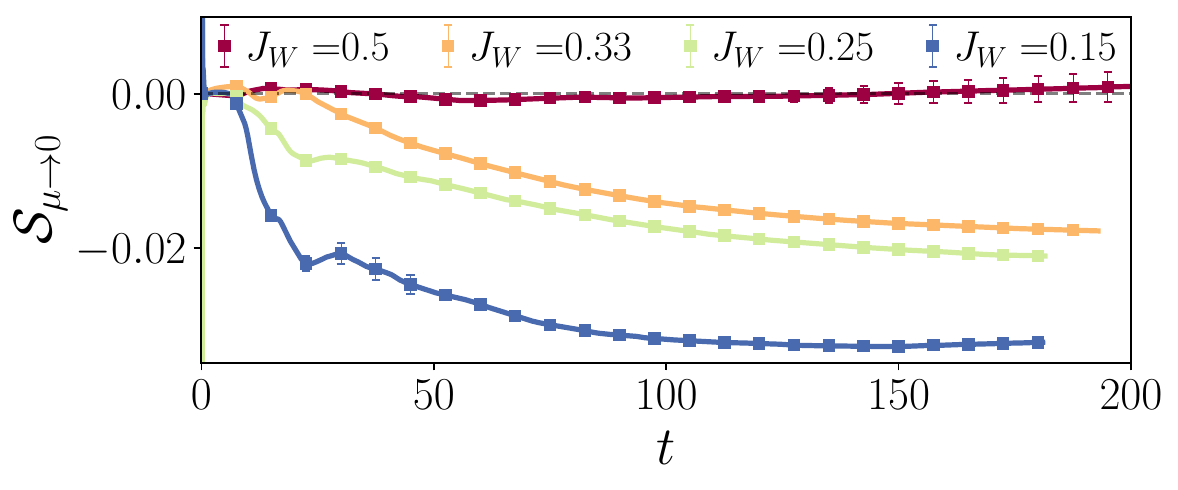}
    \caption{
    \label{Fig:skewness}
    The extrapolation of the skewness $\mathcal{S}$ for $\mu\to 0$ shows a finite long-time value for $\change{J_W}<1/2$, indicating persistent asymmetry even in the absence of the initial chemical potential step.
    The late time behavior of the skewness as a function of $\change{J_W}$ shows a monotonic decrease of the asymmetry as $\change{J_W}\to 0$.
    Due to reflection symmetry, a mirrored behavior appears with positive $\mathcal{S}$ at $\change{J_W}>1/2$.
    The data were obtained by extrapolating the intercept of a linear fit of $\mathcal{S}(t)$ vs $\mu$ for values of $\mu\in[0.001,0.05]$. The system size is $L=512$ sites, and the bond dimension used was $\chi=448$.
    }
\end{figure}

The asymmetric constraints in our Hamiltonian together with the asymmetric initial state, instead, yield rather a \textit{skewed} distribution (see~\cite{Note1} for some examples).
To quantitatively capture the amount of asymmetry in the state at a given time, we calculate the discrete skewness of the dynamical structure factor
\begin{equation}
    \label{Eq:skewness}
    \mathcal{S} = \sum_y \mathcal{P}_{\Delta n}(y) \left(\frac{y - \mu_1}{\sigma}\right)^3.
\end{equation}
Here $\mathcal{P}_{\Delta n}$ is the normalized particle density gradient, $\mu_1$ and $\sigma$ are the corresponding mean and standard deviation.

For all finite $\mu$, the skewness relaxes to a finite negative value at long times.
As we take the linear response limit $\mu\to 0$ and the asymmetry of the initial state vanishes, however, $\mathcal{S}$ is expected to vanish proportionally following the expectations for normal diffusion.
Using a linear fit, we extrapolate the skewness at $\mu=0$~\cite{Note1}, which we show in Figure~\ref{Fig:skewness}.

Surprisingly, whenever $\change{J_W}\neq \change{J_E}$ a finite skewness persists at $\mu=0$.
At the symmetric point, instead, skewness vanishes, as expected due to the symmetries in that case. 
Comparing $\mathcal{S}_{\mu\to0}$ at late times we observe a monotonic increase of the skewness as a function of $\change{J_W}$, crossing zero at the symmetric point.

While this may be expected for an asymmetric model, the \textit{asymmetric diffusive behavior} we observe deviates from expectations for diffusion.
This suggests that, while the bare transport exponent is not affected by the kinetic constraint, dynamics in general are, revealing a novel anomalous dynamical feature caused by the interplay of kinetic constraints and a lack of reflection symmetry.

\textit{Conclusions -- }%
In this work we studied the influence of kinetic constraints in combination with breaking reflection symmetry on the dynamical properties of quantum many-body systems. 
Specifically, we generalized the particle-conserving quantum East model~\cite{Brighi2023a} allowing also for West constrained hopping.
While the system exhibits chaotic level spacing statistics, we find a pure state with anomalous dynamics.
Within a range of the model parameter $\change{J_W}$, the dynamics from the LDW state exhibit several different types of transport ranging from the more typical insulating and diffusive dynamics to superdiffusion and even ballistic dynamics, typically associated to integrable systems~\cite{Hess_2001,Sologubenko_2007,Simon_2011}.
The discovery of anomalous dynamics in the LDW state invites further research into the model, specifically identifying other states sharing similar dynamics would be extremely insightful.
Furthermore, the observation of ballistic dynamics \change{in a generic state} in a chaotic model invites further research into models with broken reflection symmetry in order to understand the underlying properties that give rise to such dynamics. 

We further observed diffusive scaling at infinite temperature, consistent with the model's ergodic nature. 
However, the finite asymmetry we observed implies that the dynamics are not described by a diffusion equation with a constant diffusion coefficient. 
Indeed, due to the nature of the model, one might argue that the diffusion constant should depend on the particle density, which could explain our observations, however, the exact nature of this dependence remains an open question. 
Interestingly, the direction of the asymmetry suggests the existence of many states moving faster to the left, in contrast to the LDW state moving ballistically to the right.
These discrepancies present interesting open questions for future research, which would allow us to improve our understanding on the effects of breaking reflection symmetry on the dynamical properties of many-body quantum systems. 

Moving from dynamical to spectral properties, we observe certain anomalies in the spectrum.
Prominently, our model hosts a set of weakly entangled eigenstates, reminiscent of quantum many-body scars~\cite{Michailidis2018,Turner2018}, and an anomalously large number of zero modes only for \textit{even} particle number.
Here, similarly to the PXP model, these weakly entangled eigenstates are not engineered~\change{\cite{Mori2017}}, hence the model represents a potential avenue to improve our understanding of scars in other similar models. 

\begin{acknowledgments}
\textit{Acknowledgments -- }
The authors acknowledge useful discussions with M.~Serbyn, Z.~Papi\' c and A.~Nunnenkamp.
P.~B.~is supported by the Erwin Schr\"odinger center for Quantum Science \& Technology (ESQ) of the \"Osterreichische Akademie der Wissenschaften (\"OAW) under the  Discovery Grant.
M.~L.~acknowledges support from the European Research Council (ERC) under the European Union’s Horizon 2020 research and innovation programme (grant agreement No.~850899). 
The numerical simulations were performed using the ITensor library~\cite{ITensor} on the Vienna Scientific Cluster (VSC).
\end{acknowledgments}

\clearpage
\newpage
\onecolumngrid
\appendix
\begin{center}
    \textbf{Supplmentary material for ``\mytitle"}

    Pietro Brighi$^1$ and Marko Ljubotina$^2$

    \small{$1$ \textit{Faculty of Physics, University of Vienna, Boltzmanngasse 5, 1090 Vienna, Austria}}
    
    \small{$2$ \textit{Institute of Science and Technology Austria (ISTA), Am Campus 1, 3400, Klosterneuburg, Austria}}
\end{center}

\renewcommand{\theequation}{S\arabic{equation}}
\setcounter{equation}{0}
\renewcommand{\thefigure}{S\arabic{figure}}
\setcounter{figure}{0}
\setcounter{page}{1}
    
\section{Spectrum of the Hamiltonian, level spacing distribution and entanglement of eigenstates.}\label{App:spectrum}

\begin{figure}[b]
    \centering
    \includegraphics[width=.475\columnwidth]{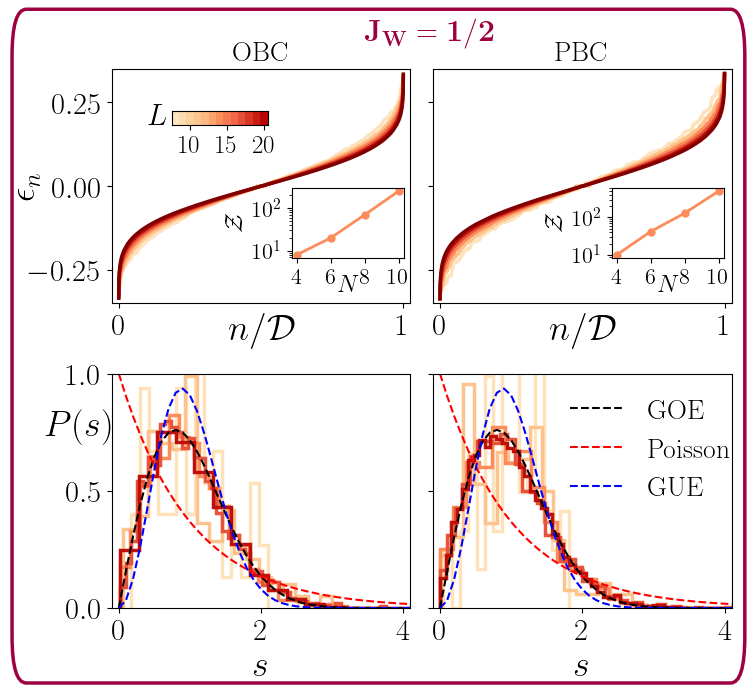} \; \includegraphics[width=.475\columnwidth]{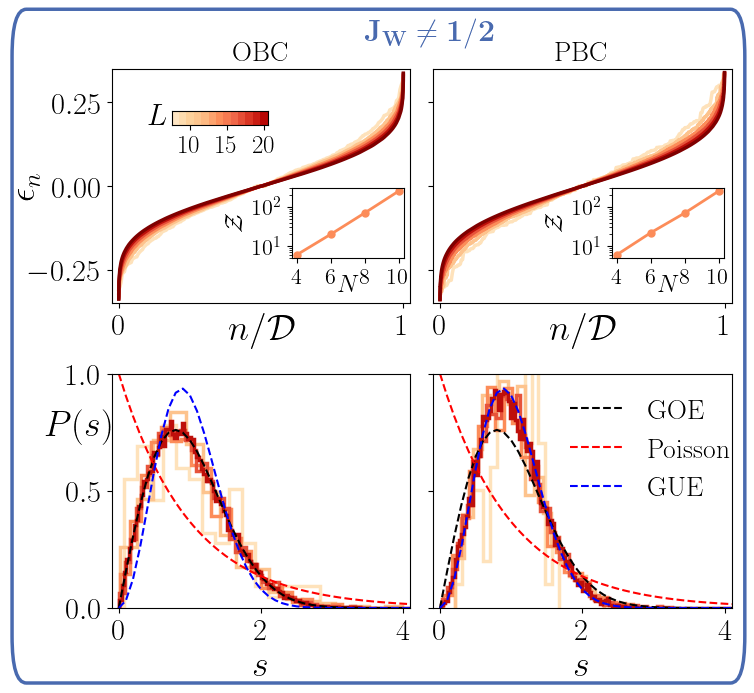}
    \caption{
    \label{Fig:Spectrum}
    In the top panels the spectra for different parameters, system sizes and boundary conditions all show the effects of \textit{particle-hole} symmetry, $E_n\to -E_n$.
    Additionally, for \textit{even} particle numbers $N$ we observe an exponential number of zero modes $\mathcal{Z}$, as shown in the insets.
    Surprisingly, these zero energy eigenstates disappear as $N$ becomes odd.
    In the bottom panels the level statistics for different parameters and boundary conditions show convergence toward chaotic behavior as the system size is increased.
    Interestingly, the PBC case with $\change{J_W}\neq1/2$ shows convergence to the Gaussian Unitary Ensemble (GUE) as opposed to the Gaussian Orthogonal Ensemble (GOE) seen in all other cases.
    }
\end{figure}

We use exact diagonalization to obtain eigenvalues and eigenstates of the Hamiltonian for system sizes up to $L=20$, always in the half-filling sector $N=L/2$.
In the case of periodic boundary conditions (PBC), we additionally resolve translational invariance, and in the case of open boundary conditions at the symmetric point $\change{J_W}=1/2$, we resolve the emergent reflection symmetry.
The spectra are shown in Figure~\ref{Fig:Spectrum} for all cases studied.
In order to show the spectra for different values of $L$, we rescale the energies as $\epsilon_n = E_n/L$ and the x-axis by the Hilbert space dimension $\mathcal{D}$.
 
\begin{figure}[t]
    \centering
    \includegraphics[width=.99\columnwidth]{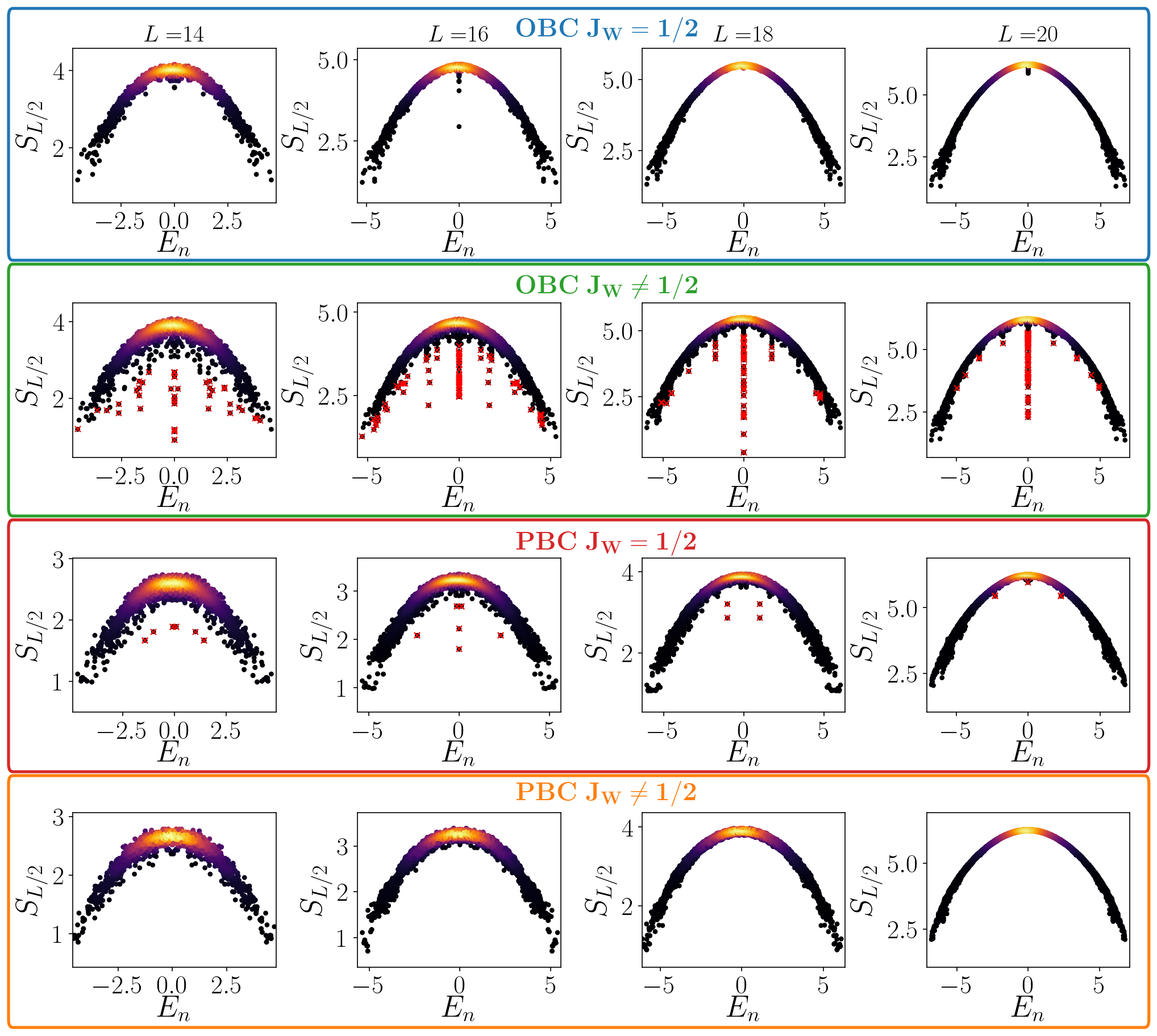}
    \caption{
    \label{Fig:Entanglement}
    Half-chain entanglement entropy of eigenstates shows anomalously weakly entangled eigenstates in the OBC case for $\change{J_W}\neq 1/2$ and at the symmetric point $\change{J_W=J_E}$ in the PBC case.
    In the case of OBC, the number of weakly entangled eigenstates grows with system size, while for PBC it seems to be reducing as $L$ increases.
    }
\end{figure}

First, we note that the Hamiltonian we study has \textit{particle-hole} symmetry, i.~e.~an operator which anti-commutes with the Hamiltonian
\begin{equation}
    \hat{\mathcal{C}}\; : \{\hat{\mathcal{C}},\hat{H}\} = 0,
\end{equation}
such that for each eigenstate with energy $E_n$, the state $\hat{\mathcal{C}}\ket{E_n}$ is also an eigenstate with opposite energy $-E_n$.
The presence of this anti-symmetry can lead to the presence of an anomalously large number of zero modes in the Hamiltonian, as every simultaneous eigenstate of $\hat{\mathcal{C}}$ and $\hat{H}$ necessarily has zero energy~\cite{Michailidis2018,Turner2018,Iadecola2018a,Papic2019}.
In our case, we do indeed notice the presence of exponentially many zero modes (the number of which we denote with $\mathcal{Z}$), as shown in the insets of Figure~\ref{Fig:Spectrum}.
Interestingly, $\mathcal{Z} \neq0$ only for even number of particles $N$, suggesting a relation with particles pairing, which might present an avenue for future research in this direction. 

Next, we focus on the level spacing distribution, i.e.~the normalized distribution of ordered nearest-neighbor eigenvalues spacing $s=E_n-E_{n-1}$.
The distribution $P(s)$, obtained through spectrum-unfolding, is a common measure of ergodicity of a model~\cite{D'Alessio2016,Oganesyan2007,Rigol2010}.
In fact, it is believed that chaotic Hamiltonians behave in line with results from Random Matrix Theory (RMT), typically showing results compatible with the Gaussian Orthogonal Ensemble (GOE) or the Gaussian Unitary Ensemble (GUE)~\cite{D'Alessio2016}
\begin{equation}
    P^\text{GOE}(s) = \frac{\pi s}{2}e^{-\frac{\pi s^2}{4}} \quad \quad P^\text{GUE}(s) = \frac{32}{\pi^2}s^2e^{-\frac{4}{\pi}s^2}.
\end{equation}
On the other hand, integrable and localized models instead exhibit a Poissonian level spacing distribution $P^\text{Poisson}(s)=\exp(-s)$.
We evaluate $P(s)$ for different system sizes and the results shown in Figure~\ref{Fig:Spectrum} clearly indicate chaotic behavior.
Interestingly, the PBC case for $\change{J_W}\neq1/2$ shows a behavior in agreement with the GUE as opposed to GOE, which is seen in all other cases.

The Gaussian Unitary Ensemble describes the level statistics of Hamiltonians with complex entries and without time-reversal symmetry.
Our Hamiltonian~(\ref{Eq:H}), instead, is fully real and therefore is expected to respect time-reversal symmetry.
As such, the presence of GUE statistics for the level spacing of the system with PBC and $\change{J_W}\neq1/2$ is unexpected and invites further investigation.

We further study the entanglement entropy of eigenstates, defined as the von Neumann entropy of the reduced density matrix obtained by tracing out half of the system
\begin{equation}
    S_{L/2} = -\tr \rho_{L/2}\log \rho_{L/2},\quad \rho_{L/2} = \tr_{i\in[1,L/2]}\rho.
\end{equation}
Highly excited eigenstates of an ergodic system are expected to have volume-law entanglement~\cite{D'Alessio2016}, which scales with the system size.
We observe such behavior for OBC with $\change{J_W}=1/2$ and for PBC with $\change{J_W}\neq1/2$.

However, as shown in Figure~\ref{Fig:Entanglement}, in the remaining cases a certain amount of weakly entangled eigenstates, reminiscent of quantum many-body scars~\cite{Turner2018}, appear.
In particular, we observe that in the OBC case at $\change{J_W}\neq1/2$, the number of such states increases with system size.
Most of the anomalous states concentrate around the center of the spectrum at $E=0$.
For even particle numbers $N$ this energy is highly degenerate, and therefore low-lying eigenstates can be obtained by mixing within the degenerate subspace~\cite{Karle}.
However, for odd $N$ our model has no exact zero modes and the weakly entangled eigenstates are close to, but not equal to, $E=0$, and are thus not degenerate.
This suggests a deeper mechanism than a simple recombination of eigenstates in the nullspace of the Hamiltonian.

\section{Convergence of tensor-network simulations}

\subsubsection{Pure Initial States}

In the main text, we used time-evolving block decimation~\cite{Vidal2003} (TEBD) to obtain the dynamics of pure initial states such as the left domain wall.
While in principle MPS methods would fail after a short time-evolution due to the rapid growth of entanglement entropy, we observe that this is not the case for the $\ket{\text{LDW}}$ state.
Indeed using bond dimensions of up to $\chi=2048$ we are able to reach times of up to $t\approx100$ for most cases. 

\begin{figure}
    \centering
    \includegraphics[width=.99\columnwidth]{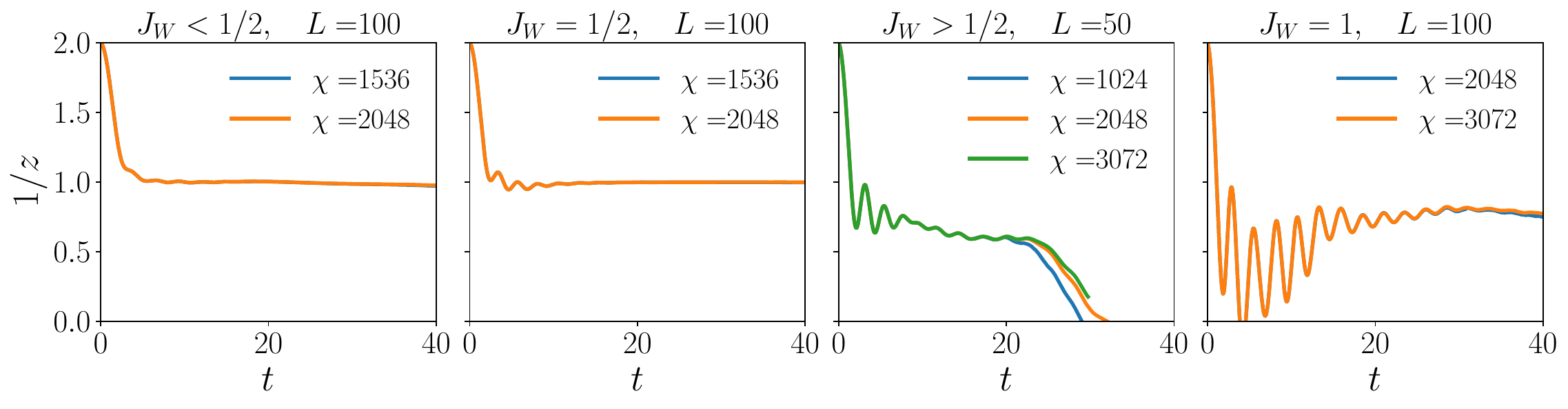}
    \caption{
    \label{Fig:bond comp states}
    Comparison of the inverse dynamical exponent for different bond dimensions shows good convergence in the $\change{J_W}\leq1/2$ case.
    Convergence is reached only at a higher bond dimension $\chi=2048$ in the $\change{J_W}=1$ case, while the hardest case, $\change{J_W}>1/2$, would require even larger bond dimensions in order to accurately simulate the dynamics up to comparable times. 
    }
\end{figure}

In Figure~\ref{Fig:bond comp states} we show the bond dimension comparison for the largest system size achieved in the different cases reported in the main text.
For $\change{J_W}\leq1/2$, where dynamics are ballistic, results are already converged at $\chi=1536$, at least up to the times we use in the main text.
The worst case corresponds to $1/2<\change{J_W}<1$ where dynamics are diffusive.
In this case, $\chi=2048$ is converged up to only $t\approx 25$, which essentially limits the system size which we are able to accurately study.
Nevertheless, our results are sufficiently accurate to demonstrate clear diffusive behavior in that regime. 
Finally, at $\change{J_W}=1$, a bond dimension of $\chi=2048$ is required to obtain converged dynamics up to the times of interest.

\subsubsection{Mixed Initial States}

\begin{figure}[t]
    \centering
    \includegraphics[width=.99\columnwidth]{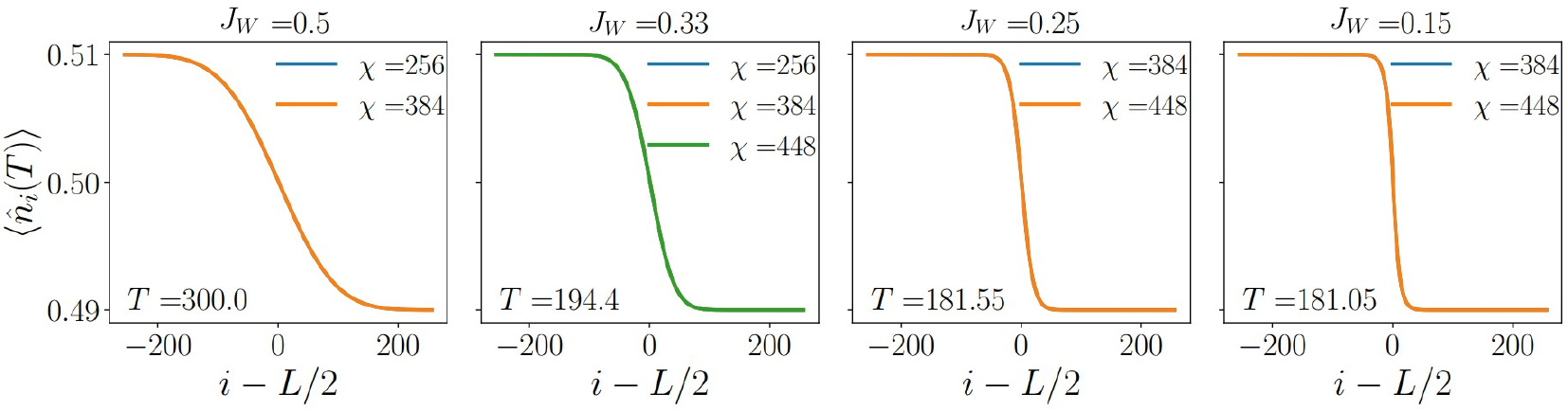}
    \caption{
    \label{Fig:Bond comparison rho}
    Comparison of different bond dimensions for $\mu=0.01$ shows excellent convergence of the density at all values of $\change{J_W}$ presented in the main text.
    }
\end{figure}

For mixed initial states such as the ones used in the second part of our work, the bond dimension typically rapidly saturates.
However, at long times, if the system relaxes towards the infinite-temperature state, a relatively small bond dimension ($\chi=256-384$) is usually sufficient to accurately capture the most significant behavior.

In Figure~\ref{Fig:Bond comparison rho}, we show the density profiles at late times for different bond dimensions and for all the values of $\change{J_W}$ shown in the main text.
We fix $\mu=0.01$, but results do not drastically change as long as $\mu\ll1$.
As one can readily observe, at late times the different curves perfectly collapse on one another, thus ensuring the convergence of our numerical simulations.

\section{Even-odd difference in dynamics}

\begin{figure}[h]
    \centering
        \includegraphics[width=.99\columnwidth]{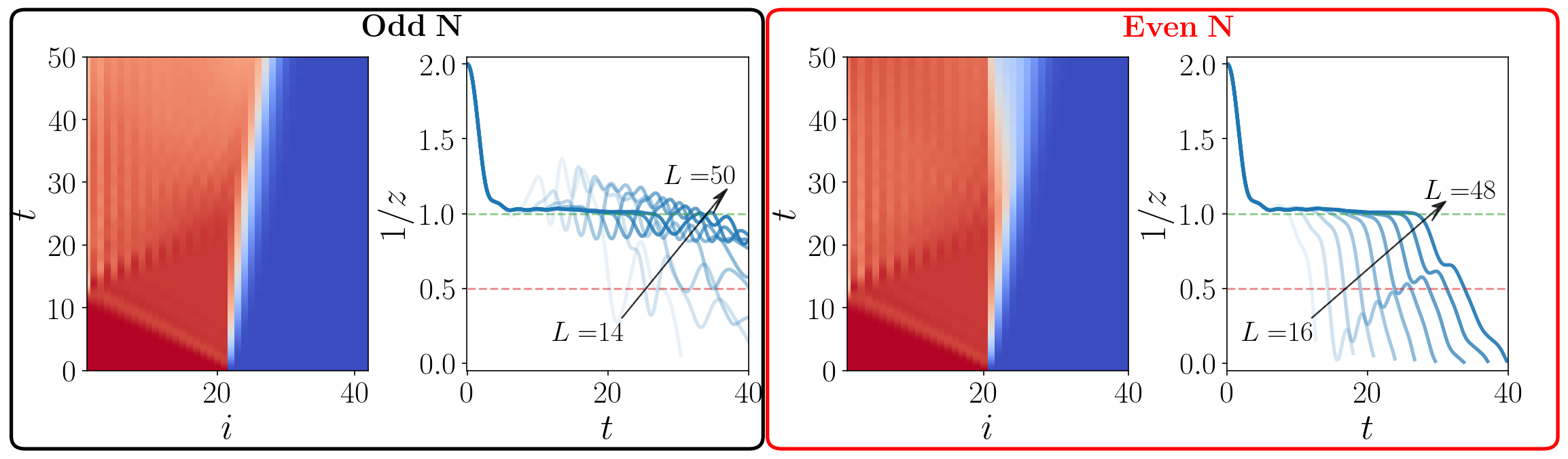}
    \caption{
    \label{Fig:even-odd dyn}
    Density dynamics when $\change{J_W}\ll \change{J_E}$ ($\change{J_W}=0.25$ in this case) shows a dramatic difference depending on the parity of the particle number.
    When $N$ is odd, particles spread ballistically through the entire chain even after finite size effects become relevant.
    However, for even $N$ particles are not able to spread past a certain point in space, and seem to localize within the left part of the chain.
    The time at which this occurs appears to coincide with the time when the boundary effects kick in. 
    }
\end{figure}

We have already observed the difference between systems with even and odd particle numbers $N$ in several properties of the model throughout this work.
Here, we show that dynamics also exhibit drastically different behavior depending on the parity of $N$ once finite size effects become relevant, especially when $\change{J_W}\ll \change{J_E}$.
In Figure~\ref{Fig:even-odd dyn}, we show the density dynamics for odd and even particle number ($L=42$ and $L=40$ respectively) for $\change{J_W} = 0.25 = \change{J_E}/3$.

In the system with odd particle number $N$, density spreads ballistically as reported in the main text even after finite size effects begin influencing the dynamics.
However, as we change to even $N$ the situation changes dramatically.
Initially, particles have the same ballistic behavior.
However, as the boundary effects kick in, around $t\approx25$, rightward spreading comes to a stop, as indicated by the quick drop of the inverse dynamical exponent.
Density in even particle systems, then, is \textit{localized} within a region $R_L = rL$, $r<1$.
While this behavior is not relevant to the dynamics in the thermodynamic limit, it is indicative of the clear differences that appear to be present in the spectral properties of the model, depending on the number of particles $N$. 

\change{

\section{Correlation functions for the domain wall state}

In this Section, we provide data for the correlation function evaluated in the domain wall initial state
\begin{equation}
    \label{Eq:n n corr}
    \langle \hat{n}_i(0) \hat{n}_j(t) \rangle= \langle\hat{n}_i(0)\rangle \langle\hat{n}_j(t)\rangle,
\end{equation}
where the equality holds as the domain wall state is an eigenstate of the density operator.
The scaling with time of the correlation functions provides a complementary way of determining the dynamical exponent of the state.
Indeed, the correlation functions rescaled by time with the appropriate exponent are expected to have a universal functional form independent of time, $\langle \hat{n}_i(0) \hat{n}_j(t) \rangle/t^{1/z} = f(i-j)$.

In Figure~\ref{Fig:sz scaling}, we report the collapse of different density profiles over the timescales available from our finite size numerical simulations.
Within the time window $t \in [25,50]$ reported here, the density profiles collapse onto one another once the space axis is rescaled by time with the dynamical exponents corresponding to the regimes reported in the main text.

\begin{figure}[h]
    \centering
    \includegraphics[width=0.99\linewidth]{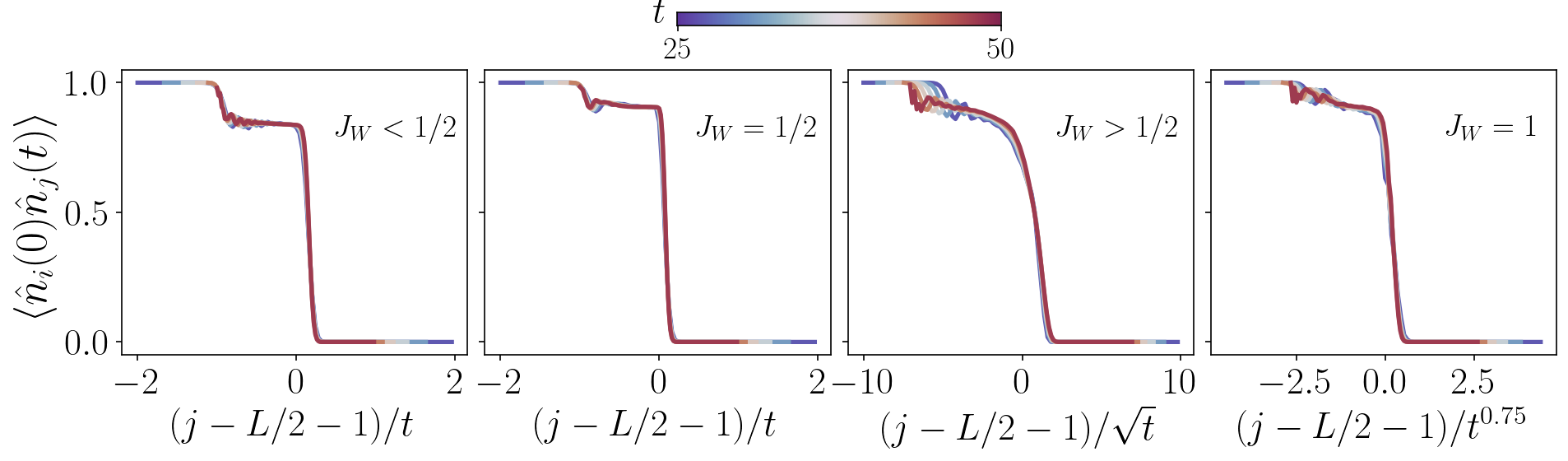}
    \caption{\label{Fig:sz scaling}
    \change{Density profiles rescaled by time with different dynamical exponents corresponding to the different regimes reported in the domain wall phase diagram in Figure~\ref{Fig:cartoon}.
    Over the range of times available $t\in [25,50]$, the different density profiles collapse onto one another, providing additional evidence for the dynamical exponents reported in the main text.
    In the case shown here, $i\leq L/2$.
    }}
\end{figure}

}

\section{Evidence of diffusion at infinite temperature}\label{App:diffusion}

In the main text, we reported evidence of diffusive transport for one particular choice of parameters of our model.
Here, we show that this feature persists in the model throughout the whole range of parameters studied.

\begin{figure}[h]
    \centering
    \includegraphics[width = 0.99\columnwidth]{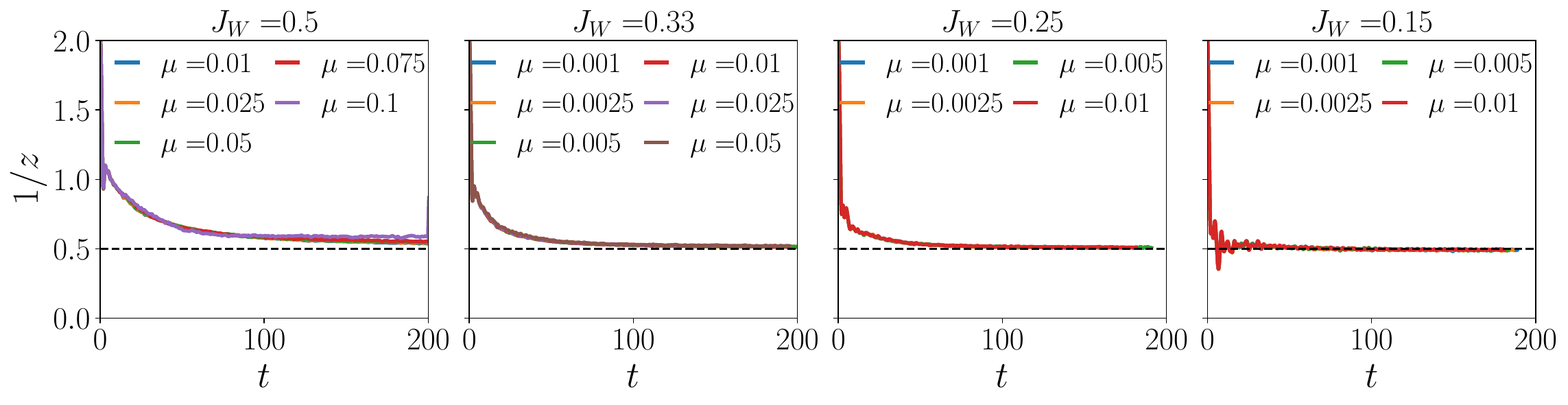}
    \caption{
    \label{Fig:z Ham TEBD}
    The dynamical exponent approaches a value compatible with diffusion irrespective of the initial step size $\mu$ both in the symmetric and asymmetric case.
    The only noticeable difference is the faster relaxation to the diffusive plateau of the asymmetric cases.
    Data are shown for $\chi = 448$ and $L=512$.
    }
\end{figure}

In Figure~\ref{Fig:z Ham TEBD}, we show the inverse dynamical exponent for many values of $\mu$ and $\change{J_W}$.
After a short transient phenomenon, $1/z$ approaches a value compatible with diffusion ($z=2$) marked by the black dashed line.
This behavior appears to be consistent irrespective of the system parameters.
As the \change{West} hopping ($\change{J_W}$) is reduced, the diffusive plateau is reached earlier in time.

\section{Details on the linear fit procedure for skewness}\label{App:skewness}

One of the most peculiar properties of our kinetically constrained model is the presence of asymmetry in the particle gradient, in spite of the diffusive scaling observed in the model.
Here, we give further details about the skewness $\mathcal{S}$ and the fitting procedure yielding Figure~\ref{Fig:skewness} in the main text.

\begin{figure}[h]
    \centering
    \includegraphics[width=.99\textwidth]{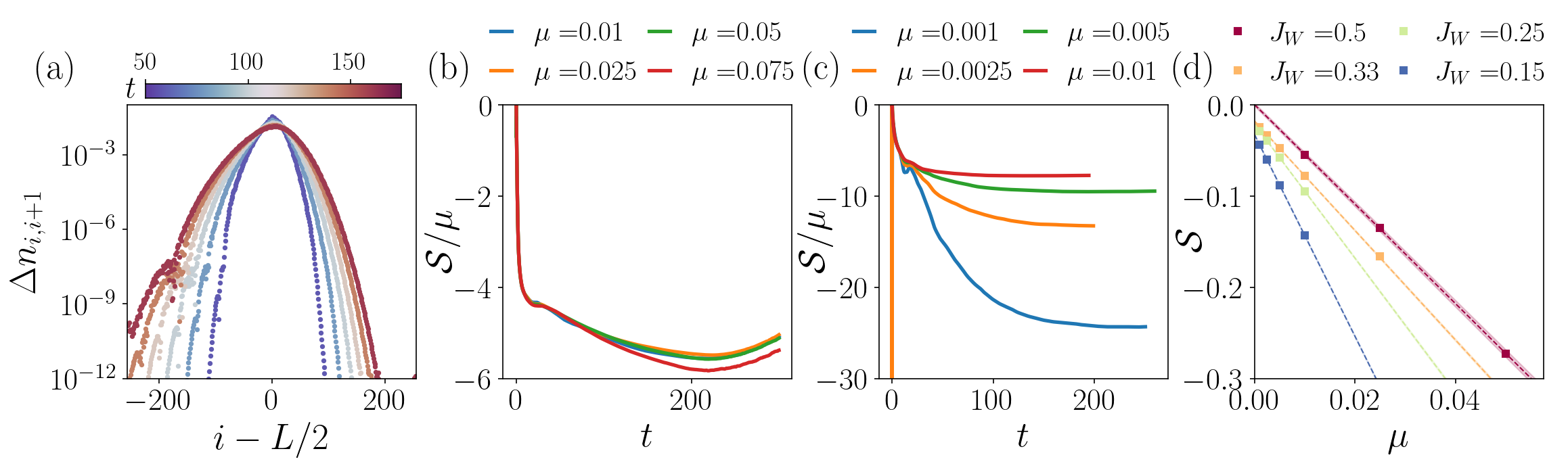}
    \caption{
    \label{Fig:Skewness SM}
    (a) Particle density gradient for $\change{J_W}=1/3$ and $\mu=0.01$ is asymmetric at all times.
    (b),(c) Skewness rescaled by the chemical potential step for $\change{J_W}=1/2$ and $\change{J_W}=1/3$ respectively.
    For $\change{J_W}=1/2$ the skewness converges to a single curve when rescaled by $\mu$. 
    On the contrary, at $\change{J_W}\neq1/2$ it diverges as $\mu\to 0$, indicating a finite value even in absence of an initial chemical potential step.
    (d) The linear fit Eq.~(\ref{Eq:f mu}) is very accurate for all values of $\change{J_W}$. The shaded area (clearly visible only for $\change{J_W}=0.5$) represents the error of the linear fit. Data are shown here for a snapshot at $t=175$.
    }
\end{figure}

First, we focus on the dynamical structure factor, expressed through the density gradient $S(i,t)=\lim_{\mu\to0}\frac 1 \mu \Delta n_{i,i+1} = \lim_{\mu\to0}\frac 1 \mu |\langle\hat{n}_i(t)\rangle_\mu - \langle\hat{n}_{i+1}(t)\rangle_\mu|$.
As mentioned in the main text, $S(i,t)$ is not symmetric around the center, which is contrary to what one would expect in a diffusive system.
In Figure~\ref{Fig:Skewness SM}(a) we show the density gradient for $\change{J_W}=1/3$ and $\mu=0.01$ over a wide range of times.
The left half of the chain shows higher values as compared with the right half, giving rise to the skewness we present in the main text.

In Figure~\ref{Fig:Skewness SM}(b) and (c) we show the skewness rescaled by the value of $\mu$ for $\change{J_W} = 1/2$ and $\change{J_W}=1/3$ respectively.
Interestingly, the two show a dramatic difference.
At the symmetric point, the curves collapse onto one another, suggesting a vanishing skewness $\mathcal{S}$ as $\mu\to0$.
On the other hand, for $\change{J_W}\neq1/2$, the rescaled skewness diverges as $\mu$ decreases, thus implying a finite value of $\mathcal{S}_0$ at $\mu=0$.

To obtain $\mathcal{S}_0$, we perform a linear fit with the function
\begin{equation}
    \label{Eq:f mu}
    f(\mu) = \mathcal{S}_0 + \alpha \mu.
\end{equation}
Examples of this fit are reported in Figure~\ref{Fig:Skewness SM}(d) at a fixed time $t=175$.
The linear fit (dashed lines) is extremely accurate, and the extracted $\mathcal{S}_0$ is non-zero whenever $\change{J_W}\neq1/2$, as reported in the main text.


\begin{thebibliography}{74}%
\makeatletter
\providecommand \@ifxundefined [1]{%
 \@ifx{#1\undefined}
}%
\providecommand \@ifnum [1]{%
 \ifnum #1\expandafter \@firstoftwo
 \else \expandafter \@secondoftwo
 \fi
}%
\providecommand \@ifx [1]{%
 \ifx #1\expandafter \@firstoftwo
 \else \expandafter \@secondoftwo
 \fi
}%
\providecommand \natexlab [1]{#1}%
\providecommand \enquote  [1]{``#1''}%
\providecommand \bibnamefont  [1]{#1}%
\providecommand \bibfnamefont [1]{#1}%
\providecommand \citenamefont [1]{#1}%
\providecommand \href@noop [0]{\@secondoftwo}%
\providecommand \href [0]{\begingroup \@sanitize@url \@href}%
\providecommand \@href[1]{\@@startlink{#1}\@@href}%
\providecommand \@@href[1]{\endgroup#1\@@endlink}%
\providecommand \@sanitize@url [0]{\catcode `\\12\catcode `\$12\catcode
  `\&12\catcode `\#12\catcode `\^12\catcode `\_12\catcode `\%12\relax}%
\providecommand \@@startlink[1]{}%
\providecommand \@@endlink[0]{}%
\providecommand \url  [0]{\begingroup\@sanitize@url \@url }%
\providecommand \@url [1]{\endgroup\@href {#1}{\urlprefix }}%
\providecommand \urlprefix  [0]{URL }%
\providecommand \Eprint [0]{\href }%
\providecommand \doibase [0]{https://doi.org/}%
\providecommand \selectlanguage [0]{\@gobble}%
\providecommand \bibinfo  [0]{\@secondoftwo}%
\providecommand \bibfield  [0]{\@secondoftwo}%
\providecommand \translation [1]{[#1]}%
\providecommand \BibitemOpen [0]{}%
\providecommand \bibitemStop [0]{}%
\providecommand \bibitemNoStop [0]{.\EOS\space}%
\providecommand \EOS [0]{\spacefactor3000\relax}%
\providecommand \BibitemShut  [1]{\csname bibitem#1\endcsname}%
\let\auto@bib@innerbib\@empty
\bibitem [{\citenamefont {Prosen}\ and\ \citenamefont
  {\ifmmode\check{Z}\else\v{Z}\fi{}nidari\ifmmode\check{c}\else\v{c}\fi{}}(2012)}]{Prosen_2012}%
  \BibitemOpen
  \bibfield  {author} {\bibinfo {author} {\bibfnamefont {T.}~\bibnamefont
  {Prosen}}\ and\ \bibinfo {author} {\bibfnamefont {M.}~\bibnamefont
  {\ifmmode\check{Z}\else\v{Z}\fi{}nidari\ifmmode\check{c}\else\v{c}\fi{}}},\
  }\bibfield  {title} {\bibinfo {title} {Diffusive high-temperature transport
  in the one-dimensional {Hubbard} model},\ }\href
  {https://doi.org/10.1103/PhysRevB.86.125118} {\bibfield  {journal} {\bibinfo
  {journal} {Phys. Rev. B}\ }\textbf {\bibinfo {volume} {86}},\ \bibinfo
  {pages} {125118} (\bibinfo {year} {2012})}\BibitemShut {NoStop}%
\bibitem [{\citenamefont {Karrasch}\ \emph {et~al.}(2014)\citenamefont
  {Karrasch}, \citenamefont {Moore},\ and\ \citenamefont
  {Heidrich-Meisner}}]{Karrasch_2014}%
  \BibitemOpen
  \bibfield  {author} {\bibinfo {author} {\bibfnamefont {C.}~\bibnamefont
  {Karrasch}}, \bibinfo {author} {\bibfnamefont {J.~E.}\ \bibnamefont
  {Moore}},\ and\ \bibinfo {author} {\bibfnamefont {F.}~\bibnamefont
  {Heidrich-Meisner}},\ }\bibfield  {title} {\bibinfo {title} {Real-time and
  real-space spin and energy dynamics in one-dimensional spin-$\frac{1}{2}$
  systems induced by local quantum quenches at finite temperatures},\ }\href
  {https://doi.org/10.1103/PhysRevB.89.075139} {\bibfield  {journal} {\bibinfo
  {journal} {Phys. Rev. B}\ }\textbf {\bibinfo {volume} {89}},\ \bibinfo
  {pages} {075139} (\bibinfo {year} {2014})}\BibitemShut {NoStop}%
\bibitem [{\citenamefont {Lux}\ \emph {et~al.}(2014)\citenamefont {Lux},
  \citenamefont {M\"uller}, \citenamefont {Mitra},\ and\ \citenamefont
  {Rosch}}]{Lux2014}%
  \BibitemOpen
  \bibfield  {author} {\bibinfo {author} {\bibfnamefont {J.}~\bibnamefont
  {Lux}}, \bibinfo {author} {\bibfnamefont {J.}~\bibnamefont {M\"uller}},
  \bibinfo {author} {\bibfnamefont {A.}~\bibnamefont {Mitra}},\ and\ \bibinfo
  {author} {\bibfnamefont {A.}~\bibnamefont {Rosch}},\ }\bibfield  {title}
  {\bibinfo {title} {Hydrodynamic long-time tails after a quantum quench},\
  }\href {https://doi.org/10.1103/PhysRevA.89.053608} {\bibfield  {journal}
  {\bibinfo  {journal} {Phys. Rev. A}\ }\textbf {\bibinfo {volume} {89}},\
  \bibinfo {pages} {053608} (\bibinfo {year} {2014})}\BibitemShut {NoStop}%
\bibitem [{\citenamefont {\ifmmode \check{Z}\else
  \v{Z}\fi{}nidari\ifmmode~\check{c}\else \v{c}\fi{}}\ \emph
  {et~al.}(2016)\citenamefont {\ifmmode \check{Z}\else
  \v{Z}\fi{}nidari\ifmmode~\check{c}\else \v{c}\fi{}}, \citenamefont
  {Scardicchio},\ and\ \citenamefont {Varma}}]{Znidaric2016}%
  \BibitemOpen
  \bibfield  {author} {\bibinfo {author} {\bibfnamefont {M.}~\bibnamefont
  {\ifmmode \check{Z}\else \v{Z}\fi{}nidari\ifmmode~\check{c}\else
  \v{c}\fi{}}}, \bibinfo {author} {\bibfnamefont {A.}~\bibnamefont
  {Scardicchio}},\ and\ \bibinfo {author} {\bibfnamefont {V.~K.}\ \bibnamefont
  {Varma}},\ }\bibfield  {title} {\bibinfo {title} {Diffusive and subdiffusive
  spin transport in the ergodic phase of a many-body localizable system},\
  }\href {https://doi.org/10.1103/PhysRevLett.117.040601} {\bibfield  {journal}
  {\bibinfo  {journal} {Phys. Rev. Lett.}\ }\textbf {\bibinfo {volume} {117}},\
  \bibinfo {pages} {040601} (\bibinfo {year} {2016})}\BibitemShut {NoStop}%
\bibitem [{\citenamefont {Gu}\ \emph {et~al.}(2017)\citenamefont {Gu},
  \citenamefont {Qi},\ and\ \citenamefont {Stanford}}]{Gu_2017}%
  \BibitemOpen
  \bibfield  {author} {\bibinfo {author} {\bibfnamefont {Y.}~\bibnamefont
  {Gu}}, \bibinfo {author} {\bibfnamefont {X.-L.}\ \bibnamefont {Qi}},\ and\
  \bibinfo {author} {\bibfnamefont {D.}~\bibnamefont {Stanford}},\ }\bibfield
  {title} {\bibinfo {title} {Local criticality, diffusion and chaos in
  generalized {Sachdev}-{Ye}-{Kitaev} models},\ }\href
  {https://doi.org/10.1007/JHEP05(2017)125} {\bibfield  {journal} {\bibinfo
  {journal} {J. High Energy Phys.}\ }\textbf {\bibinfo {volume} {2017}}\bibinfo
   {number} { (5)},\ \bibinfo {pages} {125}}\BibitemShut {NoStop}%
\bibitem [{\citenamefont {Blake}\ \emph {et~al.}(2017)\citenamefont {Blake},
  \citenamefont {Davison},\ and\ \citenamefont {Sachdev}}]{Blake_2017}%
  \BibitemOpen
\bibfield  {number} {  }\bibfield  {author} {\bibinfo {author} {\bibfnamefont
  {M.}~\bibnamefont {Blake}}, \bibinfo {author} {\bibfnamefont {R.~A.}\
  \bibnamefont {Davison}},\ and\ \bibinfo {author} {\bibfnamefont
  {S.}~\bibnamefont {Sachdev}},\ }\bibfield  {title} {\bibinfo {title} {Thermal
  diffusivity and chaos in metals without quasiparticles},\ }\href
  {https://doi.org/10.1103/PhysRevD.96.106008} {\bibfield  {journal} {\bibinfo
  {journal} {Phys. Rev. D}\ }\textbf {\bibinfo {volume} {96}},\ \bibinfo
  {pages} {106008} (\bibinfo {year} {2017})}\BibitemShut {NoStop}%
\bibitem [{\citenamefont {Friedman}\ \emph {et~al.}(2020)\citenamefont
  {Friedman}, \citenamefont {Gopalakrishnan},\ and\ \citenamefont
  {Vasseur}}]{Friedman_2020}%
  \BibitemOpen
  \bibfield  {author} {\bibinfo {author} {\bibfnamefont {A.~J.}\ \bibnamefont
  {Friedman}}, \bibinfo {author} {\bibfnamefont {S.}~\bibnamefont
  {Gopalakrishnan}},\ and\ \bibinfo {author} {\bibfnamefont {R.}~\bibnamefont
  {Vasseur}},\ }\bibfield  {title} {\bibinfo {title} {Diffusive hydrodynamics
  from integrability breaking},\ }\href
  {https://doi.org/10.1103/PhysRevB.101.180302} {\bibfield  {journal} {\bibinfo
   {journal} {Phys. Rev. B}\ }\textbf {\bibinfo {volume} {101}},\ \bibinfo
  {pages} {180302(R)} (\bibinfo {year} {2020})}\BibitemShut {NoStop}%
\bibitem [{\citenamefont {De~Nardis}\ \emph {et~al.}(2020)\citenamefont
  {De~Nardis}, \citenamefont {Medenjak}, \citenamefont {Karrasch},\ and\
  \citenamefont {Ilievski}}]{De_Nardis_2020}%
  \BibitemOpen
  \bibfield  {author} {\bibinfo {author} {\bibfnamefont {J.}~\bibnamefont
  {De~Nardis}}, \bibinfo {author} {\bibfnamefont {M.}~\bibnamefont {Medenjak}},
  \bibinfo {author} {\bibfnamefont {C.}~\bibnamefont {Karrasch}},\ and\
  \bibinfo {author} {\bibfnamefont {E.}~\bibnamefont {Ilievski}},\ }\bibfield
  {title} {\bibinfo {title} {Universality classes of spin transport in
  one-dimensional isotropic magnets: The onset of logarithmic anomalies},\
  }\href {https://doi.org/10.1103/PhysRevLett.124.210605} {\bibfield  {journal}
  {\bibinfo  {journal} {Phys. Rev. Lett.}\ }\textbf {\bibinfo {volume} {124}},\
  \bibinfo {pages} {210605} (\bibinfo {year} {2020})}\BibitemShut {NoStop}%
\bibitem [{\citenamefont {Bertini}\ \emph {et~al.}(2021)\citenamefont
  {Bertini}, \citenamefont {Heidrich-Meisner}, \citenamefont {Karrasch},
  \citenamefont {Prosen}, \citenamefont {Steinigeweg},\ and\ \citenamefont
  {\ifmmode\check{Z}\else\v{Z}\fi{}nidari\ifmmode\check{c}\else\v{c}\fi{}}}]{bertini2021finite}%
  \BibitemOpen
  \bibfield  {author} {\bibinfo {author} {\bibfnamefont {B.}~\bibnamefont
  {Bertini}}, \bibinfo {author} {\bibfnamefont {F.}~\bibnamefont
  {Heidrich-Meisner}}, \bibinfo {author} {\bibfnamefont {C.}~\bibnamefont
  {Karrasch}}, \bibinfo {author} {\bibfnamefont {T.}~\bibnamefont {Prosen}},
  \bibinfo {author} {\bibfnamefont {R.}~\bibnamefont {Steinigeweg}},\ and\
  \bibinfo {author} {\bibfnamefont {M.}~\bibnamefont
  {\ifmmode\check{Z}\else\v{Z}\fi{}nidari\ifmmode\check{c}\else\v{c}\fi{}}},\
  }\bibfield  {title} {\bibinfo {title} {Finite-temperature transport in
  one-dimensional quantum lattice models},\ }\href
  {https://doi.org/10.1103/RevModPhys.93.025003} {\bibfield  {journal}
  {\bibinfo  {journal} {Rev. Mod. Phys.}\ }\textbf {\bibinfo {volume} {93}},\
  \bibinfo {pages} {025003} (\bibinfo {year} {2021})}\BibitemShut {NoStop}%
\bibitem [{\citenamefont {Wei}\ \emph {et~al.}(2022)\citenamefont {Wei},
  \citenamefont {Rubio-Abadal}, \citenamefont {Ye}, \citenamefont {Machado},
  \citenamefont {Kemp}, \citenamefont {Srakaew}, \citenamefont {Hollerith},
  \citenamefont {Rui}, \citenamefont {Gopalakrishnan}, \citenamefont {Yao},
  \citenamefont {Bloch},\ and\ \citenamefont {Zeiher}}]{Zeiher2022}%
  \BibitemOpen
  \bibfield  {author} {\bibinfo {author} {\bibfnamefont {D.}~\bibnamefont
  {Wei}}, \bibinfo {author} {\bibfnamefont {A.}~\bibnamefont {Rubio-Abadal}},
  \bibinfo {author} {\bibfnamefont {B.}~\bibnamefont {Ye}}, \bibinfo {author}
  {\bibfnamefont {F.}~\bibnamefont {Machado}}, \bibinfo {author} {\bibfnamefont
  {J.}~\bibnamefont {Kemp}}, \bibinfo {author} {\bibfnamefont {K.}~\bibnamefont
  {Srakaew}}, \bibinfo {author} {\bibfnamefont {S.}~\bibnamefont {Hollerith}},
  \bibinfo {author} {\bibfnamefont {J.}~\bibnamefont {Rui}}, \bibinfo {author}
  {\bibfnamefont {S.}~\bibnamefont {Gopalakrishnan}}, \bibinfo {author}
  {\bibfnamefont {N.~Y.}\ \bibnamefont {Yao}}, \bibinfo {author} {\bibfnamefont
  {I.}~\bibnamefont {Bloch}},\ and\ \bibinfo {author} {\bibfnamefont
  {J.}~\bibnamefont {Zeiher}},\ }\bibfield  {title} {\bibinfo {title} {Quantum
  gas microscopy of {Kardar-Parisi-Zhang} superdiffusion},\ }\href
  {https://doi.org/10.1126/science.abk2397} {\bibfield  {journal} {\bibinfo
  {journal} {Science}\ }\textbf {\bibinfo {volume} {376}},\ \bibinfo {pages}
  {716} (\bibinfo {year} {2022})}\BibitemShut {NoStop}%
\bibitem [{\citenamefont {Basko}\ \emph {et~al.}(2006)\citenamefont {Basko},
  \citenamefont {Aleiner},\ and\ \citenamefont {Altshuler}}]{Basko06}%
  \BibitemOpen
  \bibfield  {author} {\bibinfo {author} {\bibfnamefont {D.}~\bibnamefont
  {Basko}}, \bibinfo {author} {\bibfnamefont {I.}~\bibnamefont {Aleiner}},\
  and\ \bibinfo {author} {\bibfnamefont {B.}~\bibnamefont {Altshuler}},\
  }\bibfield  {title} {\bibinfo {title} {Metal--insulator transition in a
  weakly interacting many-electron system with localized single-particle
  states},\ }\href
  {https://doi.org/http://dx.doi.org/10.1016/j.aop.2005.11.014} {\bibfield
  {journal} {\bibinfo  {journal} {Ann. Phys.}\ }\textbf {\bibinfo {volume}
  {321}},\ \bibinfo {pages} {1126 } (\bibinfo {year} {2006})}\BibitemShut
  {NoStop}%
\bibitem [{\citenamefont {Agarwal}\ \emph {et~al.}(2015)\citenamefont
  {Agarwal}, \citenamefont {Gopalakrishnan}, \citenamefont {Knap},
  \citenamefont {M\"uller},\ and\ \citenamefont {Demler}}]{Demler2015}%
  \BibitemOpen
  \bibfield  {author} {\bibinfo {author} {\bibfnamefont {K.}~\bibnamefont
  {Agarwal}}, \bibinfo {author} {\bibfnamefont {S.}~\bibnamefont
  {Gopalakrishnan}}, \bibinfo {author} {\bibfnamefont {M.}~\bibnamefont
  {Knap}}, \bibinfo {author} {\bibfnamefont {M.}~\bibnamefont {M\"uller}},\
  and\ \bibinfo {author} {\bibfnamefont {E.}~\bibnamefont {Demler}},\
  }\bibfield  {title} {\bibinfo {title} {Anomalous diffusion and {Griffiths}
  effects near the many-body localization transition},\ }\href
  {https://doi.org/10.1103/PhysRevLett.114.160401} {\bibfield  {journal}
  {\bibinfo  {journal} {Phys. Rev. Lett.}\ }\textbf {\bibinfo {volume} {114}},\
  \bibinfo {pages} {160401} (\bibinfo {year} {2015})}\BibitemShut {NoStop}%
\bibitem [{\citenamefont {Bar~Lev}\ \emph {et~al.}(2015)\citenamefont
  {Bar~Lev}, \citenamefont {Cohen},\ and\ \citenamefont
  {Reichman}}]{Reichman15}%
  \BibitemOpen
  \bibfield  {author} {\bibinfo {author} {\bibfnamefont {Y.}~\bibnamefont
  {Bar~Lev}}, \bibinfo {author} {\bibfnamefont {G.}~\bibnamefont {Cohen}},\
  and\ \bibinfo {author} {\bibfnamefont {D.~R.}\ \bibnamefont {Reichman}},\
  }\bibfield  {title} {\bibinfo {title} {Absence of diffusion in an interacting
  system of spinless fermions on a one-dimensional disordered lattice},\ }\href
  {https://doi.org/10.1103/PhysRevLett.114.100601} {\bibfield  {journal}
  {\bibinfo  {journal} {Phys. Rev. Lett.}\ }\textbf {\bibinfo {volume} {114}},\
  \bibinfo {pages} {100601} (\bibinfo {year} {2015})}\BibitemShut {NoStop}%
\bibitem [{\citenamefont {Imbrie}(2016)}]{Imbrie16}%
  \BibitemOpen
  \bibfield  {author} {\bibinfo {author} {\bibfnamefont {J.~Z.}\ \bibnamefont
  {Imbrie}},\ }\bibfield  {title} {\bibinfo {title} {On many-body localization
  for quantum spin chains},\ }\href {https://doi.org/10.1007/s10955-016-1508-x}
  {\bibfield  {journal} {\bibinfo  {journal} {Journal of Statistical Physics}\
  }\textbf {\bibinfo {volume} {163}},\ \bibinfo {pages} {998} (\bibinfo {year}
  {2016})}\BibitemShut {NoStop}%
\bibitem [{\citenamefont {Bar~Lev}\ \emph {et~al.}(2017)\citenamefont
  {Bar~Lev}, \citenamefont {Kennes}, \citenamefont {Klöckner}, \citenamefont
  {Reichman},\ and\ \citenamefont {Karrasch}}]{Bar_Lev_2017}%
  \BibitemOpen
  \bibfield  {author} {\bibinfo {author} {\bibfnamefont {Y.}~\bibnamefont
  {Bar~Lev}}, \bibinfo {author} {\bibfnamefont {D.~M.}\ \bibnamefont {Kennes}},
  \bibinfo {author} {\bibfnamefont {C.}~\bibnamefont {Klöckner}}, \bibinfo
  {author} {\bibfnamefont {D.~R.}\ \bibnamefont {Reichman}},\ and\ \bibinfo
  {author} {\bibfnamefont {C.}~\bibnamefont {Karrasch}},\ }\bibfield  {title}
  {\bibinfo {title} {Transport in quasiperiodic interacting systems: From
  superdiffusion to subdiffusion},\ }\href
  {https://doi.org/10.1209/0295-5075/119/37003} {\bibfield  {journal} {\bibinfo
   {journal} {{EPL}}\ }\textbf {\bibinfo {volume} {119}},\ \bibinfo {pages}
  {37003} (\bibinfo {year} {2017})}\BibitemShut {NoStop}%
\bibitem [{\citenamefont {{\v{Z}}nidari{\v{c}}}\ and\ \citenamefont
  {Ljubotina}(2018)}]{quasiperiodic2018}%
  \BibitemOpen
  \bibfield  {author} {\bibinfo {author} {\bibfnamefont {M.}~\bibnamefont
  {{\v{Z}}nidari{\v{c}}}}\ and\ \bibinfo {author} {\bibfnamefont
  {M.}~\bibnamefont {Ljubotina}},\ }\bibfield  {title} {\bibinfo {title}
  {Interaction instability of localization in quasiperiodic systems},\ }\href
  {https://doi.org/10.1073/pnas.1800589115} {\bibfield  {journal} {\bibinfo
  {journal} {Proc. Natl. Acad. Sci. U. S. A.}\ }\textbf {\bibinfo {volume}
  {115}},\ \bibinfo {pages} {4595} (\bibinfo {year} {2018})}\BibitemShut
  {NoStop}%
\bibitem [{\citenamefont {Ljubotina}\ \emph {et~al.}(2023)\citenamefont
  {Ljubotina}, \citenamefont {Desaules}, \citenamefont {Serbyn},\ and\
  \citenamefont {Papi\ifmmode~\acute{c}\else \'{c}\fi{}}}]{Ljubotina2023}%
  \BibitemOpen
  \bibfield  {author} {\bibinfo {author} {\bibfnamefont {M.}~\bibnamefont
  {Ljubotina}}, \bibinfo {author} {\bibfnamefont {J.-Y.}\ \bibnamefont
  {Desaules}}, \bibinfo {author} {\bibfnamefont {M.}~\bibnamefont {Serbyn}},\
  and\ \bibinfo {author} {\bibfnamefont {Z.}~\bibnamefont
  {Papi\ifmmode~\acute{c}\else \'{c}\fi{}}},\ }\bibfield  {title} {\bibinfo
  {title} {Superdiffusive energy transport in kinetically constrained models},\
  }\href {https://doi.org/10.1103/PhysRevX.13.011033} {\bibfield  {journal}
  {\bibinfo  {journal} {Phys. Rev. X}\ }\textbf {\bibinfo {volume} {13}},\
  \bibinfo {pages} {011033} (\bibinfo {year} {2023})}\BibitemShut {NoStop}%
\bibitem [{\citenamefont {Singh}\ \emph {et~al.}(2021)\citenamefont {Singh},
  \citenamefont {Ware}, \citenamefont {Vasseur},\ and\ \citenamefont
  {Friedman}}]{Vasseur2021}%
  \BibitemOpen
  \bibfield  {author} {\bibinfo {author} {\bibfnamefont {H.}~\bibnamefont
  {Singh}}, \bibinfo {author} {\bibfnamefont {B.~A.}\ \bibnamefont {Ware}},
  \bibinfo {author} {\bibfnamefont {R.}~\bibnamefont {Vasseur}},\ and\ \bibinfo
  {author} {\bibfnamefont {A.~J.}\ \bibnamefont {Friedman}},\ }\bibfield
  {title} {\bibinfo {title} {Subdiffusion and many-body quantum chaos with
  kinetic constraints},\ }\href
  {https://doi.org/10.1103/PhysRevLett.127.230602} {\bibfield  {journal}
  {\bibinfo  {journal} {Phys. Rev. Lett.}\ }\textbf {\bibinfo {volume} {127}},\
  \bibinfo {pages} {230602} (\bibinfo {year} {2021})}\BibitemShut {NoStop}%
\bibitem [{\citenamefont {van Horssen}\ \emph {et~al.}(2015)\citenamefont {van
  Horssen}, \citenamefont {Levi},\ and\ \citenamefont
  {Garrahan}}]{Garrahan2015}%
  \BibitemOpen
  \bibfield  {author} {\bibinfo {author} {\bibfnamefont {M.}~\bibnamefont {van
  Horssen}}, \bibinfo {author} {\bibfnamefont {E.}~\bibnamefont {Levi}},\ and\
  \bibinfo {author} {\bibfnamefont {J.~P.}\ \bibnamefont {Garrahan}},\
  }\bibfield  {title} {\bibinfo {title} {Dynamics of many-body localization in
  a translation-invariant quantum glass model},\ }\href
  {https://doi.org/10.1103/PhysRevB.92.100305} {\bibfield  {journal} {\bibinfo
  {journal} {Phys. Rev. B}\ }\textbf {\bibinfo {volume} {92}},\ \bibinfo
  {pages} {100305(R)} (\bibinfo {year} {2015})}\BibitemShut {NoStop}%
\bibitem [{\citenamefont {Lan}\ \emph {et~al.}(2018)\citenamefont {Lan},
  \citenamefont {van Horssen}, \citenamefont {Powell},\ and\ \citenamefont
  {Garrahan}}]{Garrahan2018}%
  \BibitemOpen
  \bibfield  {author} {\bibinfo {author} {\bibfnamefont {Z.}~\bibnamefont
  {Lan}}, \bibinfo {author} {\bibfnamefont {M.}~\bibnamefont {van Horssen}},
  \bibinfo {author} {\bibfnamefont {S.}~\bibnamefont {Powell}},\ and\ \bibinfo
  {author} {\bibfnamefont {J.~P.}\ \bibnamefont {Garrahan}},\ }\bibfield
  {title} {\bibinfo {title} {Quantum slow relaxation and metastability due to
  dynamical constraints},\ }\href
  {https://doi.org/10.1103/PhysRevLett.121.040603} {\bibfield  {journal}
  {\bibinfo  {journal} {Phys. Rev. Lett.}\ }\textbf {\bibinfo {volume} {121}},\
  \bibinfo {pages} {040603} (\bibinfo {year} {2018})}\BibitemShut {NoStop}%
\bibitem [{\citenamefont {Turner}\ \emph
  {et~al.}(2018{\natexlab{a}})\citenamefont {Turner}, \citenamefont
  {Michailidis}, \citenamefont {Abanin}, \citenamefont {Serbyn},\ and\
  \citenamefont {Papi\ifmmode~\acute{c}\else \'{c}\fi{}}}]{Michailidis2018}%
  \BibitemOpen
  \bibfield  {author} {\bibinfo {author} {\bibfnamefont {C.~J.}\ \bibnamefont
  {Turner}}, \bibinfo {author} {\bibfnamefont {A.~A.}\ \bibnamefont
  {Michailidis}}, \bibinfo {author} {\bibfnamefont {D.~A.}\ \bibnamefont
  {Abanin}}, \bibinfo {author} {\bibfnamefont {M.}~\bibnamefont {Serbyn}},\
  and\ \bibinfo {author} {\bibfnamefont {Z.}~\bibnamefont
  {Papi\ifmmode~\acute{c}\else \'{c}\fi{}}},\ }\bibfield  {title} {\bibinfo
  {title} {Weak ergodicity breaking from quantum many-body scars},\ }\href
  {https://doi.org/10.1038/s41567-018-0137-5} {\bibfield  {journal} {\bibinfo
  {journal} {Nat. Phys.}\ }\textbf {\bibinfo {volume} {14}},\ \bibinfo {pages}
  {745} (\bibinfo {year} {2018}{\natexlab{a}})}\BibitemShut {NoStop}%
\bibitem [{\citenamefont {Pancotti}\ \emph {et~al.}(2020)\citenamefont
  {Pancotti}, \citenamefont {Giudice}, \citenamefont {Cirac}, \citenamefont
  {Garrahan},\ and\ \citenamefont {Ba\~nuls}}]{Pancotti2020}%
  \BibitemOpen
  \bibfield  {author} {\bibinfo {author} {\bibfnamefont {N.}~\bibnamefont
  {Pancotti}}, \bibinfo {author} {\bibfnamefont {G.}~\bibnamefont {Giudice}},
  \bibinfo {author} {\bibfnamefont {J.~I.}\ \bibnamefont {Cirac}}, \bibinfo
  {author} {\bibfnamefont {J.~P.}\ \bibnamefont {Garrahan}},\ and\ \bibinfo
  {author} {\bibfnamefont {M.~C.}\ \bibnamefont {Ba\~nuls}},\ }\bibfield
  {title} {\bibinfo {title} {Quantum {East} model: Localization, nonthermal
  eigenstates, and slow dynamics},\ }\href
  {https://doi.org/10.1103/PhysRevX.10.021051} {\bibfield  {journal} {\bibinfo
  {journal} {Phys. Rev. X}\ }\textbf {\bibinfo {volume} {10}},\ \bibinfo
  {pages} {021051} (\bibinfo {year} {2020})}\BibitemShut {NoStop}%
\bibitem [{\citenamefont {Valencia-Tortora}\ \emph {et~al.}(2022)\citenamefont
  {Valencia-Tortora}, \citenamefont {Pancotti},\ and\ \citenamefont
  {Marino}}]{Marino2022}%
  \BibitemOpen
  \bibfield  {author} {\bibinfo {author} {\bibfnamefont {R.~J.}\ \bibnamefont
  {Valencia-Tortora}}, \bibinfo {author} {\bibfnamefont {N.}~\bibnamefont
  {Pancotti}},\ and\ \bibinfo {author} {\bibfnamefont {J.}~\bibnamefont
  {Marino}},\ }\bibfield  {title} {\bibinfo {title} {Kinetically constrained
  quantum dynamics in superconducting circuits},\ }\href
  {https://doi.org/10.1103/PRXQuantum.3.020346} {\bibfield  {journal} {\bibinfo
   {journal} {PRX Quantum}\ }\textbf {\bibinfo {volume} {3}},\ \bibinfo {pages}
  {020346} (\bibinfo {year} {2022})}\BibitemShut {NoStop}%
\bibitem [{\citenamefont {Brighi}\ \emph {et~al.}(2023)\citenamefont {Brighi},
  \citenamefont {Ljubotina},\ and\ \citenamefont {Serbyn}}]{Brighi2023a}%
  \BibitemOpen
  \bibfield  {author} {\bibinfo {author} {\bibfnamefont {P.}~\bibnamefont
  {Brighi}}, \bibinfo {author} {\bibfnamefont {M.}~\bibnamefont {Ljubotina}},\
  and\ \bibinfo {author} {\bibfnamefont {M.}~\bibnamefont {Serbyn}},\
  }\bibfield  {title} {\bibinfo {title} {{Hilbert space fragmentation and slow
  dynamics in particle-conserving quantum East models}},\ }\href
  {https://doi.org/10.21468/SciPostPhys.15.3.093} {\bibfield  {journal}
  {\bibinfo  {journal} {SciPost Phys.}\ }\textbf {\bibinfo {volume} {15}},\
  \bibinfo {pages} {093} (\bibinfo {year} {2023})}\BibitemShut {NoStop}%
\bibitem [{\citenamefont {Bernien}\ \emph {et~al.}(2017)\citenamefont
  {Bernien}, \citenamefont {Schwartz}, \citenamefont {Keesling}, \citenamefont
  {Levine}, \citenamefont {Omran}, \citenamefont {Pichler}, \citenamefont
  {Choi}, \citenamefont {Zibrov}, \citenamefont {Endres}, \citenamefont
  {Greiner}, \citenamefont {Vuletić},\ and\ \citenamefont
  {Lukin}}]{Lukin2017}%
  \BibitemOpen
  \bibfield  {author} {\bibinfo {author} {\bibfnamefont {H.}~\bibnamefont
  {Bernien}}, \bibinfo {author} {\bibfnamefont {S.}~\bibnamefont {Schwartz}},
  \bibinfo {author} {\bibfnamefont {A.}~\bibnamefont {Keesling}}, \bibinfo
  {author} {\bibfnamefont {H.}~\bibnamefont {Levine}}, \bibinfo {author}
  {\bibfnamefont {A.}~\bibnamefont {Omran}}, \bibinfo {author} {\bibfnamefont
  {H.}~\bibnamefont {Pichler}}, \bibinfo {author} {\bibfnamefont
  {S.}~\bibnamefont {Choi}}, \bibinfo {author} {\bibfnamefont {A.~S.}\
  \bibnamefont {Zibrov}}, \bibinfo {author} {\bibfnamefont {M.}~\bibnamefont
  {Endres}}, \bibinfo {author} {\bibfnamefont {M.}~\bibnamefont {Greiner}},
  \bibinfo {author} {\bibfnamefont {V.}~\bibnamefont {Vuletić}},\ and\
  \bibinfo {author} {\bibfnamefont {M.~D.}\ \bibnamefont {Lukin}},\ }\bibfield
  {title} {\bibinfo {title} {Probing many-body dynamics on a 51-atom quantum
  simulator},\ }\href {https://doi.org/10.1038/nature24622} {\bibfield
  {journal} {\bibinfo  {journal} {Nature}\ }\textbf {\bibinfo {volume} {551}},\
  \bibinfo {pages} {579} (\bibinfo {year} {2017})}\BibitemShut {NoStop}%
\bibitem [{\citenamefont {Fredrickson}\ and\ \citenamefont
  {Andersen}(1984)}]{Andersen1984}%
  \BibitemOpen
  \bibfield  {author} {\bibinfo {author} {\bibfnamefont {G.~H.}\ \bibnamefont
  {Fredrickson}}\ and\ \bibinfo {author} {\bibfnamefont {H.~C.}\ \bibnamefont
  {Andersen}},\ }\bibfield  {title} {\bibinfo {title} {Kinetic {Ising} model of
  the glass transition},\ }\href {https://doi.org/10.1103/PhysRevLett.53.1244}
  {\bibfield  {journal} {\bibinfo  {journal} {Phys. Rev. Lett.}\ }\textbf
  {\bibinfo {volume} {53}},\ \bibinfo {pages} {1244} (\bibinfo {year}
  {1984})}\BibitemShut {NoStop}%
\bibitem [{\citenamefont {Kob}\ and\ \citenamefont
  {Andersen}(1993)}]{Andersen1993}%
  \BibitemOpen
  \bibfield  {author} {\bibinfo {author} {\bibfnamefont {W.}~\bibnamefont
  {Kob}}\ and\ \bibinfo {author} {\bibfnamefont {H.~C.}\ \bibnamefont
  {Andersen}},\ }\bibfield  {title} {\bibinfo {title} {Kinetic lattice-gas
  model of cage effects in high-density liquids and a test of mode-coupling
  theory of the ideal-glass transition},\ }\href
  {https://doi.org/10.1103/PhysRevE.48.4364} {\bibfield  {journal} {\bibinfo
  {journal} {Phys. Rev. E}\ }\textbf {\bibinfo {volume} {48}},\ \bibinfo
  {pages} {4364} (\bibinfo {year} {1993})}\BibitemShut {NoStop}%
\bibitem [{\citenamefont {Ritort}\ and\ \citenamefont
  {Sollich}(2003)}]{Ritort2003}%
  \BibitemOpen
  \bibfield  {author} {\bibinfo {author} {\bibfnamefont {F.}~\bibnamefont
  {Ritort}}\ and\ \bibinfo {author} {\bibfnamefont {P.}~\bibnamefont
  {Sollich}},\ }\bibfield  {title} {\bibinfo {title} {Glassy dynamics of
  kinetically constrained models},\ }\href
  {https://doi.org/10.1080/0001873031000093582} {\bibfield  {journal} {\bibinfo
   {journal} {Adv. Phys.}\ }\textbf {\bibinfo {volume} {52}},\ \bibinfo {pages}
  {219} (\bibinfo {year} {2003})}\BibitemShut {NoStop}%
\bibitem [{\citenamefont {Rakovszky}\ \emph {et~al.}(2020)\citenamefont
  {Rakovszky}, \citenamefont {Sala}, \citenamefont {Verresen}, \citenamefont
  {Knap},\ and\ \citenamefont {Pollmann}}]{Pollmann2020a}%
  \BibitemOpen
  \bibfield  {author} {\bibinfo {author} {\bibfnamefont {T.}~\bibnamefont
  {Rakovszky}}, \bibinfo {author} {\bibfnamefont {P.}~\bibnamefont {Sala}},
  \bibinfo {author} {\bibfnamefont {R.}~\bibnamefont {Verresen}}, \bibinfo
  {author} {\bibfnamefont {M.}~\bibnamefont {Knap}},\ and\ \bibinfo {author}
  {\bibfnamefont {F.}~\bibnamefont {Pollmann}},\ }\bibfield  {title} {\bibinfo
  {title} {Statistical localization: From strong fragmentation to strong edge
  modes},\ }\href {https://doi.org/10.1103/PhysRevB.101.125126} {\bibfield
  {journal} {\bibinfo  {journal} {Phys. Rev. B}\ }\textbf {\bibinfo {volume}
  {101}},\ \bibinfo {pages} {125126} (\bibinfo {year} {2020})}\BibitemShut
  {NoStop}%
\bibitem [{\citenamefont {Khemani}\ \emph {et~al.}(2020)\citenamefont
  {Khemani}, \citenamefont {Hermele},\ and\ \citenamefont
  {Nandkishore}}]{Nandkishore2020}%
  \BibitemOpen
  \bibfield  {author} {\bibinfo {author} {\bibfnamefont {V.}~\bibnamefont
  {Khemani}}, \bibinfo {author} {\bibfnamefont {M.}~\bibnamefont {Hermele}},\
  and\ \bibinfo {author} {\bibfnamefont {R.}~\bibnamefont {Nandkishore}},\
  }\bibfield  {title} {\bibinfo {title} {Localization from {Hilbert} space
  shattering: From theory to physical realizations},\ }\href
  {https://doi.org/10.1103/PhysRevB.101.174204} {\bibfield  {journal} {\bibinfo
   {journal} {Phys. Rev. B}\ }\textbf {\bibinfo {volume} {101}},\ \bibinfo
  {pages} {174204} (\bibinfo {year} {2020})}\BibitemShut {NoStop}%
\bibitem [{\citenamefont {Yang}\ \emph {et~al.}(2020)\citenamefont {Yang},
  \citenamefont {Liu}, \citenamefont {Gorshkov},\ and\ \citenamefont
  {Iadecola}}]{Iadecola2020}%
  \BibitemOpen
  \bibfield  {author} {\bibinfo {author} {\bibfnamefont {Z.-C.}\ \bibnamefont
  {Yang}}, \bibinfo {author} {\bibfnamefont {F.}~\bibnamefont {Liu}}, \bibinfo
  {author} {\bibfnamefont {A.~V.}\ \bibnamefont {Gorshkov}},\ and\ \bibinfo
  {author} {\bibfnamefont {T.}~\bibnamefont {Iadecola}},\ }\bibfield  {title}
  {\bibinfo {title} {Hilbert-space fragmentation from strict confinement},\
  }\href {https://doi.org/10.1103/PhysRevLett.124.207602} {\bibfield  {journal}
  {\bibinfo  {journal} {Phys. Rev. Lett.}\ }\textbf {\bibinfo {volume} {124}},\
  \bibinfo {pages} {207602} (\bibinfo {year} {2020})}\BibitemShut {NoStop}%
\bibitem [{\citenamefont {Pozsgay}\ \emph {et~al.}(2021)\citenamefont
  {Pozsgay}, \citenamefont {Gombor}, \citenamefont {Hutsalyuk}, \citenamefont
  {Jiang}, \citenamefont {Pristy\'ak},\ and\ \citenamefont
  {Vernier}}]{Pozsgay2021}%
  \BibitemOpen
  \bibfield  {author} {\bibinfo {author} {\bibfnamefont {B.}~\bibnamefont
  {Pozsgay}}, \bibinfo {author} {\bibfnamefont {T.}~\bibnamefont {Gombor}},
  \bibinfo {author} {\bibfnamefont {A.}~\bibnamefont {Hutsalyuk}}, \bibinfo
  {author} {\bibfnamefont {Y.}~\bibnamefont {Jiang}}, \bibinfo {author}
  {\bibfnamefont {L.}~\bibnamefont {Pristy\'ak}},\ and\ \bibinfo {author}
  {\bibfnamefont {E.}~\bibnamefont {Vernier}},\ }\bibfield  {title} {\bibinfo
  {title} {Integrable spin chain with {Hilbert} space fragmentation and
  solvable real-time dynamics},\ }\href
  {https://doi.org/10.1103/PhysRevE.104.044106} {\bibfield  {journal} {\bibinfo
   {journal} {Phys. Rev. E}\ }\textbf {\bibinfo {volume} {104}},\ \bibinfo
  {pages} {044106} (\bibinfo {year} {2021})}\BibitemShut {NoStop}%
\bibitem [{\citenamefont {Mukherjee}\ \emph {et~al.}(2021)\citenamefont
  {Mukherjee}, \citenamefont {Banerjee}, \citenamefont {Sengupta},\ and\
  \citenamefont {Sen}}]{Sen2021}%
  \BibitemOpen
  \bibfield  {author} {\bibinfo {author} {\bibfnamefont {B.}~\bibnamefont
  {Mukherjee}}, \bibinfo {author} {\bibfnamefont {D.}~\bibnamefont {Banerjee}},
  \bibinfo {author} {\bibfnamefont {K.}~\bibnamefont {Sengupta}},\ and\
  \bibinfo {author} {\bibfnamefont {A.}~\bibnamefont {Sen}},\ }\bibfield
  {title} {\bibinfo {title} {Minimal model for {Hilbert} space fragmentation
  with local constraints},\ }\href
  {https://doi.org/10.1103/PhysRevB.104.155117} {\bibfield  {journal} {\bibinfo
   {journal} {Phys. Rev. B}\ }\textbf {\bibinfo {volume} {104}},\ \bibinfo
  {pages} {155117} (\bibinfo {year} {2021})}\BibitemShut {NoStop}%
\bibitem [{\citenamefont {Zadnik}\ and\ \citenamefont
  {Fagotti}(2021)}]{Zadnik2021}%
  \BibitemOpen
  \bibfield  {author} {\bibinfo {author} {\bibfnamefont {L.}~\bibnamefont
  {Zadnik}}\ and\ \bibinfo {author} {\bibfnamefont {M.}~\bibnamefont
  {Fagotti}},\ }\bibfield  {title} {\bibinfo {title} {{The Folded Spin-1/2 XXZ
  Model: I. Diagonalisation, Jamming, and Ground State Properties}},\ }\href
  {https://doi.org/10.21468/SciPostPhysCore.4.2.010} {\bibfield  {journal}
  {\bibinfo  {journal} {SciPost Phys. Core}\ }\textbf {\bibinfo {volume} {4}},\
  \bibinfo {pages} {010} (\bibinfo {year} {2021})}\BibitemShut {NoStop}%
\bibitem [{\citenamefont {Moudgalya}\ and\ \citenamefont
  {Motrunich}(2022)}]{Motrunic2022}%
  \BibitemOpen
  \bibfield  {author} {\bibinfo {author} {\bibfnamefont {S.}~\bibnamefont
  {Moudgalya}}\ and\ \bibinfo {author} {\bibfnamefont {O.~I.}\ \bibnamefont
  {Motrunich}},\ }\bibfield  {title} {\bibinfo {title} {Hilbert space
  fragmentation and commutant algebras},\ }\href
  {https://doi.org/10.1103/PhysRevX.12.011050} {\bibfield  {journal} {\bibinfo
  {journal} {Phys. Rev. X}\ }\textbf {\bibinfo {volume} {12}},\ \bibinfo
  {pages} {011050} (\bibinfo {year} {2022})}\BibitemShut {NoStop}%
\bibitem [{\citenamefont {Turner}\ \emph
  {et~al.}(2018{\natexlab{b}})\citenamefont {Turner}, \citenamefont
  {Michailidis}, \citenamefont {Abanin}, \citenamefont {Serbyn},\ and\
  \citenamefont {Papi\ifmmode~\acute{c}\else \'{c}\fi{}}}]{Turner2018}%
  \BibitemOpen
  \bibfield  {author} {\bibinfo {author} {\bibfnamefont {C.~J.}\ \bibnamefont
  {Turner}}, \bibinfo {author} {\bibfnamefont {A.~A.}\ \bibnamefont
  {Michailidis}}, \bibinfo {author} {\bibfnamefont {D.~A.}\ \bibnamefont
  {Abanin}}, \bibinfo {author} {\bibfnamefont {M.}~\bibnamefont {Serbyn}},\
  and\ \bibinfo {author} {\bibfnamefont {Z.}~\bibnamefont
  {Papi\ifmmode~\acute{c}\else \'{c}\fi{}}},\ }\bibfield  {title} {\bibinfo
  {title} {Quantum scarred eigenstates in a {Rydberg} atom chain: Entanglement,
  breakdown of thermalization, and stability to perturbations},\ }\href
  {https://doi.org/10.1103/PhysRevB.98.155134} {\bibfield  {journal} {\bibinfo
  {journal} {Phys. Rev. B}\ }\textbf {\bibinfo {volume} {98}},\ \bibinfo
  {pages} {155134} (\bibinfo {year} {2018}{\natexlab{b}})}\BibitemShut
  {NoStop}%
\bibitem [{\citenamefont {Moudgalya}\ \emph {et~al.}(2018)\citenamefont
  {Moudgalya}, \citenamefont {Regnault},\ and\ \citenamefont
  {Bernevig}}]{Bernevig2018}%
  \BibitemOpen
  \bibfield  {author} {\bibinfo {author} {\bibfnamefont {S.}~\bibnamefont
  {Moudgalya}}, \bibinfo {author} {\bibfnamefont {N.}~\bibnamefont
  {Regnault}},\ and\ \bibinfo {author} {\bibfnamefont {B.~A.}\ \bibnamefont
  {Bernevig}},\ }\bibfield  {title} {\bibinfo {title} {Entanglement of exact
  excited states of {Affleck-Kennedy-Lieb-Tasaki} models: Exact results,
  many-body scars, and violation of the strong eigenstate thermalization
  hypothesis},\ }\href {https://doi.org/10.1103/PhysRevB.98.235156} {\bibfield
  {journal} {\bibinfo  {journal} {Phys. Rev. B}\ }\textbf {\bibinfo {volume}
  {98}},\ \bibinfo {pages} {235156} (\bibinfo {year} {2018})}\BibitemShut
  {NoStop}%
\bibitem [{\citenamefont {Choi}\ \emph {et~al.}(2019)\citenamefont {Choi},
  \citenamefont {Turner}, \citenamefont {Pichler}, \citenamefont {Ho},
  \citenamefont {Michailidis}, \citenamefont {Papi\ifmmode~\acute{c}\else
  \'{c}\fi{}}, \citenamefont {Serbyn}, \citenamefont {Lukin},\ and\
  \citenamefont {Abanin}}]{Choi2019}%
  \BibitemOpen
  \bibfield  {author} {\bibinfo {author} {\bibfnamefont {S.}~\bibnamefont
  {Choi}}, \bibinfo {author} {\bibfnamefont {C.~J.}\ \bibnamefont {Turner}},
  \bibinfo {author} {\bibfnamefont {H.}~\bibnamefont {Pichler}}, \bibinfo
  {author} {\bibfnamefont {W.~W.}\ \bibnamefont {Ho}}, \bibinfo {author}
  {\bibfnamefont {A.~A.}\ \bibnamefont {Michailidis}}, \bibinfo {author}
  {\bibfnamefont {Z.}~\bibnamefont {Papi\ifmmode~\acute{c}\else \'{c}\fi{}}},
  \bibinfo {author} {\bibfnamefont {M.}~\bibnamefont {Serbyn}}, \bibinfo
  {author} {\bibfnamefont {M.~D.}\ \bibnamefont {Lukin}},\ and\ \bibinfo
  {author} {\bibfnamefont {D.~A.}\ \bibnamefont {Abanin}},\ }\bibfield  {title}
  {\bibinfo {title} {Emergent su(2) dynamics and perfect quantum many-body
  scars},\ }\href {https://doi.org/10.1103/PhysRevLett.122.220603} {\bibfield
  {journal} {\bibinfo  {journal} {Phys. Rev. Lett.}\ }\textbf {\bibinfo
  {volume} {122}},\ \bibinfo {pages} {220603} (\bibinfo {year}
  {2019})}\BibitemShut {NoStop}%
\bibitem [{\citenamefont {Iadecola}\ \emph {et~al.}(2019)\citenamefont
  {Iadecola}, \citenamefont {Schecter},\ and\ \citenamefont
  {Xu}}]{Iadecola2019}%
  \BibitemOpen
  \bibfield  {author} {\bibinfo {author} {\bibfnamefont {T.}~\bibnamefont
  {Iadecola}}, \bibinfo {author} {\bibfnamefont {M.}~\bibnamefont {Schecter}},\
  and\ \bibinfo {author} {\bibfnamefont {S.}~\bibnamefont {Xu}},\ }\bibfield
  {title} {\bibinfo {title} {Quantum many-body scars from magnon
  condensation},\ }\href {https://doi.org/10.1103/PhysRevB.100.184312}
  {\bibfield  {journal} {\bibinfo  {journal} {Phys. Rev. B}\ }\textbf {\bibinfo
  {volume} {100}},\ \bibinfo {pages} {184312} (\bibinfo {year}
  {2019})}\BibitemShut {NoStop}%
\bibitem [{\citenamefont {Ho}\ \emph {et~al.}(2019)\citenamefont {Ho},
  \citenamefont {Choi}, \citenamefont {Pichler},\ and\ \citenamefont
  {Lukin}}]{Lukin2019}%
  \BibitemOpen
  \bibfield  {author} {\bibinfo {author} {\bibfnamefont {W.~W.}\ \bibnamefont
  {Ho}}, \bibinfo {author} {\bibfnamefont {S.}~\bibnamefont {Choi}}, \bibinfo
  {author} {\bibfnamefont {H.}~\bibnamefont {Pichler}},\ and\ \bibinfo {author}
  {\bibfnamefont {M.~D.}\ \bibnamefont {Lukin}},\ }\bibfield  {title} {\bibinfo
  {title} {Periodic orbits, entanglement, and quantum many-body scars in
  constrained models: Matrix product state approach},\ }\href
  {https://doi.org/10.1103/PhysRevLett.122.040603} {\bibfield  {journal}
  {\bibinfo  {journal} {Phys. Rev. Lett.}\ }\textbf {\bibinfo {volume} {122}},\
  \bibinfo {pages} {040603} (\bibinfo {year} {2019})}\BibitemShut {NoStop}%
\bibitem [{\citenamefont {Hudomal}\ \emph {et~al.}(2020)\citenamefont
  {Hudomal}, \citenamefont {Vasi\ifmmode~\acute{c}\else \'{c}\fi{}},
  \citenamefont {Regnault},\ and\ \citenamefont {Papi\ifmmode~\acute{c}\else
  \'{c}\fi{}}}]{Papic2020}%
  \BibitemOpen
  \bibfield  {author} {\bibinfo {author} {\bibfnamefont {A.}~\bibnamefont
  {Hudomal}}, \bibinfo {author} {\bibfnamefont {I.}~\bibnamefont
  {Vasi\ifmmode~\acute{c}\else \'{c}\fi{}}}, \bibinfo {author} {\bibfnamefont
  {N.}~\bibnamefont {Regnault}},\ and\ \bibinfo {author} {\bibfnamefont
  {Z.}~\bibnamefont {Papi\ifmmode~\acute{c}\else \'{c}\fi{}}},\ }\bibfield
  {title} {\bibinfo {title} {Quantum scars of bosons with correlated hopping},\
  }\href {https://doi.org/10.1038/s42005-020-0364-9} {\bibfield  {journal}
  {\bibinfo  {journal} {Comm. Phys.}\ }\textbf {\bibinfo {volume} {3}},\
  \bibinfo {pages} {99} (\bibinfo {year} {2020})}\BibitemShut {NoStop}%
\bibitem [{\citenamefont {Zhao}\ \emph {et~al.}(2020)\citenamefont {Zhao},
  \citenamefont {Vovrosh}, \citenamefont {Mintert},\ and\ \citenamefont
  {Knolle}}]{Knolle2020}%
  \BibitemOpen
  \bibfield  {author} {\bibinfo {author} {\bibfnamefont {H.}~\bibnamefont
  {Zhao}}, \bibinfo {author} {\bibfnamefont {J.}~\bibnamefont {Vovrosh}},
  \bibinfo {author} {\bibfnamefont {F.}~\bibnamefont {Mintert}},\ and\ \bibinfo
  {author} {\bibfnamefont {J.}~\bibnamefont {Knolle}},\ }\bibfield  {title}
  {\bibinfo {title} {Quantum many-body scars in optical lattices},\ }\href
  {https://doi.org/10.1103/PhysRevLett.124.160604} {\bibfield  {journal}
  {\bibinfo  {journal} {Phys. Rev. Lett.}\ }\textbf {\bibinfo {volume} {124}},\
  \bibinfo {pages} {160604} (\bibinfo {year} {2020})}\BibitemShut {NoStop}%
\bibitem [{\citenamefont {Zhao}\ \emph {et~al.}(2021)\citenamefont {Zhao},
  \citenamefont {Smith}, \citenamefont {Mintert},\ and\ \citenamefont
  {Knolle}}]{Knolle2021}%
  \BibitemOpen
  \bibfield  {author} {\bibinfo {author} {\bibfnamefont {H.}~\bibnamefont
  {Zhao}}, \bibinfo {author} {\bibfnamefont {A.}~\bibnamefont {Smith}},
  \bibinfo {author} {\bibfnamefont {F.}~\bibnamefont {Mintert}},\ and\ \bibinfo
  {author} {\bibfnamefont {J.}~\bibnamefont {Knolle}},\ }\bibfield  {title}
  {\bibinfo {title} {Orthogonal quantum many-body scars},\ }\href
  {https://doi.org/10.1103/PhysRevLett.127.150601} {\bibfield  {journal}
  {\bibinfo  {journal} {Phys. Rev. Lett.}\ }\textbf {\bibinfo {volume} {127}},\
  \bibinfo {pages} {150601} (\bibinfo {year} {2021})}\BibitemShut {NoStop}%
\bibitem [{\citenamefont {Tamura}\ and\ \citenamefont
  {Katsura}(2022)}]{Tamura2022}%
  \BibitemOpen
  \bibfield  {author} {\bibinfo {author} {\bibfnamefont {K.}~\bibnamefont
  {Tamura}}\ and\ \bibinfo {author} {\bibfnamefont {H.}~\bibnamefont
  {Katsura}},\ }\bibfield  {title} {\bibinfo {title} {Quantum many-body scars
  of spinless fermions with density-assisted hopping in higher dimensions},\
  }\href {https://doi.org/10.1103/PhysRevB.106.144306} {\bibfield  {journal}
  {\bibinfo  {journal} {Phys. Rev. B}\ }\textbf {\bibinfo {volume} {106}},\
  \bibinfo {pages} {144306} (\bibinfo {year} {2022})}\BibitemShut {NoStop}%
\bibitem [{\citenamefont {Moudgalya}\ \emph {et~al.}(2022)\citenamefont
  {Moudgalya}, \citenamefont {Bernevig},\ and\ \citenamefont
  {Regnault}}]{Regnault2022}%
  \BibitemOpen
  \bibfield  {author} {\bibinfo {author} {\bibfnamefont {S.}~\bibnamefont
  {Moudgalya}}, \bibinfo {author} {\bibfnamefont {B.~A.}\ \bibnamefont
  {Bernevig}},\ and\ \bibinfo {author} {\bibfnamefont {N.}~\bibnamefont
  {Regnault}},\ }\bibfield  {title} {\bibinfo {title} {Quantum many-body scars
  and {Hilbert} space fragmentation: a review of exact results},\ }\href
  {https://doi.org/10.1088/1361-6633/ac73a0} {\bibfield  {journal} {\bibinfo
  {journal} {Rep. Prog. Phys.}\ }\textbf {\bibinfo {volume} {85}},\ \bibinfo
  {pages} {086501} (\bibinfo {year} {2022})}\BibitemShut {NoStop}%
\bibitem [{\citenamefont {Serbyn}\ \emph {et~al.}(2021)\citenamefont {Serbyn},
  \citenamefont {Abanin},\ and\ \citenamefont {Papi{\'c}}}]{Serbyn2021}%
  \BibitemOpen
  \bibfield  {author} {\bibinfo {author} {\bibfnamefont {M.}~\bibnamefont
  {Serbyn}}, \bibinfo {author} {\bibfnamefont {D.~A.}\ \bibnamefont {Abanin}},\
  and\ \bibinfo {author} {\bibfnamefont {Z.}~\bibnamefont {Papi{\'c}}},\
  }\bibfield  {title} {\bibinfo {title} {Quantum many-body scars and weak
  breaking of ergodicity},\ }\href {https://doi.org/10.1038/s41567-021-01230-2}
  {\bibfield  {journal} {\bibinfo  {journal} {Nat. Phys.}\ }\textbf {\bibinfo
  {volume} {17}},\ \bibinfo {pages} {675} (\bibinfo {year} {2021})}\BibitemShut
  {NoStop}%
\bibitem [{\citenamefont {Gei\ss{}ler}\ and\ \citenamefont
  {Garrahan}(2023)}]{Garrahan2023}%
  \BibitemOpen
  \bibfield  {author} {\bibinfo {author} {\bibfnamefont {A.}~\bibnamefont
  {Gei\ss{}ler}}\ and\ \bibinfo {author} {\bibfnamefont {J.~P.}\ \bibnamefont
  {Garrahan}},\ }\bibfield  {title} {\bibinfo {title} {Slow dynamics and
  nonergodicity of the bosonic quantum {East} model in the semiclassical
  limit},\ }\href {https://doi.org/10.1103/PhysRevE.108.034207} {\bibfield
  {journal} {\bibinfo  {journal} {Phys. Rev. E}\ }\textbf {\bibinfo {volume}
  {108}},\ \bibinfo {pages} {034207} (\bibinfo {year} {2023})}\BibitemShut
  {NoStop}%
\bibitem [{\citenamefont {Bertini}\ \emph
  {et~al.}(2024{\natexlab{a}})\citenamefont {Bertini}, \citenamefont {Kos},\
  and\ \citenamefont {Prosen}}]{Prosen2023}%
  \BibitemOpen
  \bibfield  {author} {\bibinfo {author} {\bibfnamefont {B.}~\bibnamefont
  {Bertini}}, \bibinfo {author} {\bibfnamefont {P.}~\bibnamefont {Kos}},\ and\
  \bibinfo {author} {\bibfnamefont {T.}~\bibnamefont {Prosen}},\ }\bibfield
  {title} {\bibinfo {title} {Localized dynamics in the {Floquet} quantum {East}
  model},\ }\href {https://doi.org/10.1103/PhysRevLett.132.080401} {\bibfield
  {journal} {\bibinfo  {journal} {Phys. Rev. Lett.}\ }\textbf {\bibinfo
  {volume} {132}},\ \bibinfo {pages} {080401} (\bibinfo {year}
  {2024}{\natexlab{a}})}\BibitemShut {NoStop}%
\bibitem [{\citenamefont {Bertini}\ \emph
  {et~al.}(2024{\natexlab{b}})\citenamefont {Bertini}, \citenamefont
  {De~Fazio}, \citenamefont {Garrahan},\ and\ \citenamefont
  {Klobas}}]{Garrahan2023a}%
  \BibitemOpen
  \bibfield  {author} {\bibinfo {author} {\bibfnamefont {B.}~\bibnamefont
  {Bertini}}, \bibinfo {author} {\bibfnamefont {C.}~\bibnamefont {De~Fazio}},
  \bibinfo {author} {\bibfnamefont {J.~P.}\ \bibnamefont {Garrahan}},\ and\
  \bibinfo {author} {\bibfnamefont {K.}~\bibnamefont {Klobas}},\ }\bibfield
  {title} {\bibinfo {title} {Exact quench dynamics of the {Floquet} quantum
  east model at the deterministic point},\ }\href
  {https://doi.org/10.1103/physrevlett.132.120402} {\bibfield  {journal}
  {\bibinfo  {journal} {Phys. Rev. Lett.}\ }\textbf {\bibinfo {volume} {132}},\
  \bibinfo {pages} {120402} (\bibinfo {year} {2024}{\natexlab{b}})}\BibitemShut
  {NoStop}%
\bibitem [{Note1()}]{Note1}%
  \BibitemOpen
  \bibinfo {note} {\label {footnoteSM}See supplementary material for weakly
  entangled eigenstates, anomalous zero modes, convergence analysis and
  additional information about the skewness fitting procedure.}\BibitemShut
  {Stop}%
\bibitem [{\citenamefont {Antal}\ \emph {et~al.}(1999)\citenamefont {Antal},
  \citenamefont {R{\'a}cz}, \citenamefont {R{á}kos},\ and\ \citenamefont
  {Sch{ü}tz}}]{Antal_1999}%
  \BibitemOpen
  \bibfield  {author} {\bibinfo {author} {\bibfnamefont {T.}~\bibnamefont
  {Antal}}, \bibinfo {author} {\bibfnamefont {Z.}~\bibnamefont {R{\'a}cz}},
  \bibinfo {author} {\bibfnamefont {A.}~\bibnamefont {R{á}kos}},\ and\
  \bibinfo {author} {\bibfnamefont {G.~M.}\ \bibnamefont {Sch{ü}tz}},\
  }\bibfield  {title} {\bibinfo {title} {Transport in the {$XX$} chain at zero
  temperature: Emergence of flat magnetization profiles},\ }\href
  {https://doi.org/10.1103/physreve.59.4912} {\bibfield  {journal} {\bibinfo
  {journal} {Phys. Rev. E}\ }\textbf {\bibinfo {volume} {59}},\ \bibinfo
  {pages} {4912–4918} (\bibinfo {year} {1999})}\BibitemShut {NoStop}%
\bibitem [{\citenamefont {Gobert}\ \emph {et~al.}(2005)\citenamefont {Gobert},
  \citenamefont {Kollath}, \citenamefont {{Schollwöck}},\ and\ \citenamefont
  {{Schütz}}}]{Gobert_2005}%
  \BibitemOpen
  \bibfield  {author} {\bibinfo {author} {\bibfnamefont {D.}~\bibnamefont
  {Gobert}}, \bibinfo {author} {\bibfnamefont {C.}~\bibnamefont {Kollath}},
  \bibinfo {author} {\bibfnamefont {U.}~\bibnamefont {{Schollwöck}}},\ and\
  \bibinfo {author} {\bibfnamefont {G.}~\bibnamefont {{Schütz}}},\ }\bibfield
  {title} {\bibinfo {title} {Real-time dynamics in spin-1/2 chains with
  adaptive time-dependent density matrix renormalization group},\ }\href
  {https://doi.org/10.1103/physreve.71.036102} {\bibfield  {journal} {\bibinfo
  {journal} {Phys. Rev. E}\ }\textbf {\bibinfo {volume} {71}},\ \bibinfo
  {pages} {036102} (\bibinfo {year} {2005})}\BibitemShut {NoStop}%
\bibitem [{\citenamefont {Jiang}\ and\ \citenamefont
  {Devereaux}(2019)}]{Jiang_2019}%
  \BibitemOpen
  \bibfield  {author} {\bibinfo {author} {\bibfnamefont {H.-C.}\ \bibnamefont
  {Jiang}}\ and\ \bibinfo {author} {\bibfnamefont {T.~P.}\ \bibnamefont
  {Devereaux}},\ }\bibfield  {title} {\bibinfo {title} {Superconductivity in
  the doped hubbard model and its interplay with next-nearest hopping t{'}},\
  }\href {https://doi.org/10.1126/science.aal5304} {\bibfield  {journal}
  {\bibinfo  {journal} {Science}\ }\textbf {\bibinfo {volume} {365}},\ \bibinfo
  {pages} {1424–1428} (\bibinfo {year} {2019})}\BibitemShut {NoStop}%
\bibitem [{\citenamefont {Dong}\ \emph {et~al.}(2022)\citenamefont {Dong},
  \citenamefont {Del~Re}, \citenamefont {Toschi},\ and\ \citenamefont
  {Gull}}]{Dong_2022}%
  \BibitemOpen
  \bibfield  {author} {\bibinfo {author} {\bibfnamefont {X.}~\bibnamefont
  {Dong}}, \bibinfo {author} {\bibfnamefont {L.}~\bibnamefont {Del~Re}},
  \bibinfo {author} {\bibfnamefont {A.}~\bibnamefont {Toschi}},\ and\ \bibinfo
  {author} {\bibfnamefont {E.}~\bibnamefont {Gull}},\ }\bibfield  {title}
  {\bibinfo {title} {Mechanism of superconductivity in the hubbard model at
  intermediate interaction strength},\ }\href
  {https://doi.org/10.1073/pnas.2205048119} {\bibfield  {journal} {\bibinfo
  {journal} {Proc. Natl. Acad. Sci. U. S. A.}\ }\textbf {\bibinfo {volume}
  {119}},\ \bibinfo {pages} {e2205048119} (\bibinfo {year} {2022})}\BibitemShut
  {NoStop}%
\bibitem [{\citenamefont {Ljubotina}\ \emph {et~al.}(2019)\citenamefont
  {Ljubotina}, \citenamefont {\ifmmode \check{Z}\else
  \v{Z}\fi{}nidari\ifmmode~\check{c}\else \v{c}\fi{}},\ and\ \citenamefont
  {Prosen}}]{Prosen2019}%
  \BibitemOpen
  \bibfield  {author} {\bibinfo {author} {\bibfnamefont {M.}~\bibnamefont
  {Ljubotina}}, \bibinfo {author} {\bibfnamefont {M.}~\bibnamefont {\ifmmode
  \check{Z}\else \v{Z}\fi{}nidari\ifmmode~\check{c}\else \v{c}\fi{}}},\ and\
  \bibinfo {author} {\bibfnamefont {T.}~\bibnamefont {Prosen}},\ }\bibfield
  {title} {\bibinfo {title} {{Kardar-Parisi-Zhang} physics in the quantum
  {Heisenberg} magnet},\ }\href
  {https://doi.org/10.1103/PhysRevLett.122.210602} {\bibfield  {journal}
  {\bibinfo  {journal} {Phys. Rev. Lett.}\ }\textbf {\bibinfo {volume} {122}},\
  \bibinfo {pages} {210602} (\bibinfo {year} {2019})}\BibitemShut {NoStop}%
\bibitem [{\citenamefont {Vidal}(2003)}]{Vidal2003}%
  \BibitemOpen
  \bibfield  {author} {\bibinfo {author} {\bibfnamefont {G.}~\bibnamefont
  {Vidal}},\ }\bibfield  {title} {\bibinfo {title} {Efficient classical
  simulation of slightly entangled quantum computations},\ }\href
  {https://doi.org/10.1103/PhysRevLett.91.147902} {\bibfield  {journal}
  {\bibinfo  {journal} {Phys. Rev. Lett.}\ }\textbf {\bibinfo {volume} {91}},\
  \bibinfo {pages} {147902} (\bibinfo {year} {2003})}\BibitemShut {NoStop}%
\bibitem [{Note2()}]{Note2}%
  \BibitemOpen
  \bibinfo {note} {Note that the LDW state is an
  eigenstate of $\protect \hat {n}_i(0)$, thus computing the expectation value
  of $\protect \hat {n}_j(t)$ gives us direct access to the correlation
  function $\protect \hat {n}_i(0)\protect \hat {n}_j(t)$ evaluated in the LDW
  state. See supplementary material for further details}\BibitemShut {NoStop}%
\bibitem [{\citenamefont {Sutherland}(2004)}]{Sutherland2004}%
  \BibitemOpen
  \bibfield  {author} {\bibinfo {author} {\bibfnamefont {B.}~\bibnamefont
  {Sutherland}},\ }\href@noop {} {\emph {\bibinfo {title} {Beautiful models: 70
  years of exactly solved quantum many-body problems}}}\ (\bibinfo  {publisher}
  {World Scientific},\ \bibinfo {year} {2004})\BibitemShut {NoStop}%
\bibitem [{\citenamefont {Ilievski}\ \emph {et~al.}(2021)\citenamefont
  {Ilievski}, \citenamefont {De~Nardis}, \citenamefont {Gopalakrishnan},
  \citenamefont {Vasseur},\ and\ \citenamefont {Ware}}]{Ilievski2021}%
  \BibitemOpen
  \bibfield  {author} {\bibinfo {author} {\bibfnamefont {E.}~\bibnamefont
  {Ilievski}}, \bibinfo {author} {\bibfnamefont {J.}~\bibnamefont {De~Nardis}},
  \bibinfo {author} {\bibfnamefont {S.}~\bibnamefont {Gopalakrishnan}},
  \bibinfo {author} {\bibfnamefont {R.}~\bibnamefont {Vasseur}},\ and\ \bibinfo
  {author} {\bibfnamefont {B.}~\bibnamefont {Ware}},\ }\bibfield  {title}
  {\bibinfo {title} {Superuniversality of superdiffusion},\ }\href
  {https://doi.org/10.1103/PhysRevX.11.031023} {\bibfield  {journal} {\bibinfo
  {journal} {Phys. Rev. X}\ }\textbf {\bibinfo {volume} {11}},\ \bibinfo
  {pages} {031023} (\bibinfo {year} {2021})}\BibitemShut {NoStop}%
\bibitem [{\citenamefont {\ifmmode \check{Z}\else
  \v{Z}\fi{}nidari\ifmmode~\check{c}\else \v{c}\fi{}}(2011)}]{Znidaric2011}%
  \BibitemOpen
  \bibfield  {author} {\bibinfo {author} {\bibfnamefont {M.}~\bibnamefont
  {\ifmmode \check{Z}\else \v{Z}\fi{}nidari\ifmmode~\check{c}\else
  \v{c}\fi{}}},\ }\bibfield  {title} {\bibinfo {title} {Spin transport in a
  one-dimensional anisotropic {Heisenberg} model},\ }\href
  {https://doi.org/10.1103/PhysRevLett.106.220601} {\bibfield  {journal}
  {\bibinfo  {journal} {Phys. Rev. Lett.}\ }\textbf {\bibinfo {volume} {106}},\
  \bibinfo {pages} {220601} (\bibinfo {year} {2011})}\BibitemShut {NoStop}%
\bibitem [{\citenamefont {Ljubotina}\ \emph {et~al.}(2017)\citenamefont
  {Ljubotina}, \citenamefont {{\v Z}nidari{\v c}},\ and\ \citenamefont
  {Prosen}}]{Ljubotina2017}%
  \BibitemOpen
  \bibfield  {author} {\bibinfo {author} {\bibfnamefont {M.}~\bibnamefont
  {Ljubotina}}, \bibinfo {author} {\bibfnamefont {M.}~\bibnamefont {{\v
  Z}nidari{\v c}}},\ and\ \bibinfo {author} {\bibfnamefont {T.}~\bibnamefont
  {Prosen}},\ }\bibfield  {title} {\bibinfo {title} {Spin diffusion from an
  inhomogeneous quench in an integrable system},\ }\href
  {https://doi.org/10.1038/ncomms16117} {\bibfield  {journal} {\bibinfo
  {journal} {Nat. Commun.}\ }\textbf {\bibinfo {volume} {8}},\ \bibinfo {pages}
  {16117} (\bibinfo {year} {2017})}\BibitemShut {NoStop}%
\bibitem [{\citenamefont {Ilievski}\ \emph {et~al.}(2018)\citenamefont
  {Ilievski}, \citenamefont {De~Nardis}, \citenamefont {Medenjak},\ and\
  \citenamefont {Prosen}}]{Prosen2018}%
  \BibitemOpen
  \bibfield  {author} {\bibinfo {author} {\bibfnamefont {E.}~\bibnamefont
  {Ilievski}}, \bibinfo {author} {\bibfnamefont {J.}~\bibnamefont {De~Nardis}},
  \bibinfo {author} {\bibfnamefont {M.}~\bibnamefont {Medenjak}},\ and\
  \bibinfo {author} {\bibfnamefont {T.}~\bibnamefont {Prosen}},\ }\bibfield
  {title} {\bibinfo {title} {Superdiffusion in one-dimensional quantum lattice
  models},\ }\href {https://doi.org/10.1103/PhysRevLett.121.230602} {\bibfield
  {journal} {\bibinfo  {journal} {Phys. Rev. Lett.}\ }\textbf {\bibinfo
  {volume} {121}},\ \bibinfo {pages} {230602} (\bibinfo {year}
  {2018})}\BibitemShut {NoStop}%
\bibitem [{\citenamefont {Iadecola}\ and\ \citenamefont
  {\v{Z}nidari\v{c}}(2019)}]{Znidaric2019}%
  \BibitemOpen
  \bibfield  {author} {\bibinfo {author} {\bibfnamefont {T.}~\bibnamefont
  {Iadecola}}\ and\ \bibinfo {author} {\bibfnamefont {M.}~\bibnamefont
  {\v{Z}nidari\v{c}}},\ }\bibfield  {title} {\bibinfo {title} {Exact localized
  and ballistic eigenstates in disordered chaotic spin ladders and the
  fermi-hubbard model},\ }\href
  {https://doi.org/10.1103/PhysRevLett.123.036403} {\bibfield  {journal}
  {\bibinfo  {journal} {Phys. Rev. Lett.}\ }\textbf {\bibinfo {volume} {123}},\
  \bibinfo {pages} {036403} (\bibinfo {year} {2019})}\BibitemShut {NoStop}%
\bibitem [{\citenamefont {Hess}\ \emph {et~al.}(2001)\citenamefont {Hess},
  \citenamefont {Baumann}, \citenamefont {Ammerahl}, \citenamefont
  {{Büchner}}, \citenamefont {Heidrich-Meisner}, \citenamefont {Brenig},\ and\
  \citenamefont {Revcolevschi}}]{Hess_2001}%
  \BibitemOpen
  \bibfield  {author} {\bibinfo {author} {\bibfnamefont {C.}~\bibnamefont
  {Hess}}, \bibinfo {author} {\bibfnamefont {C.}~\bibnamefont {Baumann}},
  \bibinfo {author} {\bibfnamefont {U.}~\bibnamefont {Ammerahl}}, \bibinfo
  {author} {\bibfnamefont {B.}~\bibnamefont {{Büchner}}}, \bibinfo {author}
  {\bibfnamefont {F.}~\bibnamefont {Heidrich-Meisner}}, \bibinfo {author}
  {\bibfnamefont {W.}~\bibnamefont {Brenig}},\ and\ \bibinfo {author}
  {\bibfnamefont {A.}~\bibnamefont {Revcolevschi}},\ }\bibfield  {title}
  {\bibinfo {title} {Magnon heat transport in
  {$(\mathrm{Sr},\mathrm{Ca},\mathrm{La}{)}_{14}{\mathrm{Cu}}_{24}{\mathrm{O}}_{41}$}},\
  }\href {https://doi.org/10.1103/physrevb.64.184305} {\bibfield  {journal}
  {\bibinfo  {journal} {Phys. Rev. B}\ }\textbf {\bibinfo {volume} {64}},\
  \bibinfo {pages} {184305} (\bibinfo {year} {2001})}\BibitemShut {NoStop}%
\bibitem [{\citenamefont {Sologubenko}\ \emph {et~al.}(2007)\citenamefont
  {Sologubenko}, \citenamefont {Lorenz}, \citenamefont {Ott},\ and\
  \citenamefont {Freimuth}}]{Sologubenko_2007}%
  \BibitemOpen
  \bibfield  {author} {\bibinfo {author} {\bibfnamefont {A.~V.}\ \bibnamefont
  {Sologubenko}}, \bibinfo {author} {\bibfnamefont {T.}~\bibnamefont {Lorenz}},
  \bibinfo {author} {\bibfnamefont {H.~R.}\ \bibnamefont {Ott}},\ and\ \bibinfo
  {author} {\bibfnamefont {A.}~\bibnamefont {Freimuth}},\ }\bibfield  {title}
  {\bibinfo {title} {Thermal conductivity via magnetic excitations in
  spin-chain materials},\ }\href {https://doi.org/10.1007/s10909-007-9317-x}
  {\bibfield  {journal} {\bibinfo  {journal} {J. Low Temp. Phys.}\ }\textbf
  {\bibinfo {volume} {147}},\ \bibinfo {pages} {387–403} (\bibinfo {year}
  {2007})}\BibitemShut {NoStop}%
\bibitem [{\citenamefont {Simon}\ \emph {et~al.}(2011)\citenamefont {Simon},
  \citenamefont {Bakr}, \citenamefont {Ma}, \citenamefont {Tai}, \citenamefont
  {Preiss},\ and\ \citenamefont {Greiner}}]{Simon_2011}%
  \BibitemOpen
  \bibfield  {author} {\bibinfo {author} {\bibfnamefont {J.}~\bibnamefont
  {Simon}}, \bibinfo {author} {\bibfnamefont {W.~S.}\ \bibnamefont {Bakr}},
  \bibinfo {author} {\bibfnamefont {R.}~\bibnamefont {Ma}}, \bibinfo {author}
  {\bibfnamefont {M.~E.}\ \bibnamefont {Tai}}, \bibinfo {author} {\bibfnamefont
  {P.~M.}\ \bibnamefont {Preiss}},\ and\ \bibinfo {author} {\bibfnamefont
  {M.}~\bibnamefont {Greiner}},\ }\bibfield  {title} {\bibinfo {title} {Quantum
  simulation of antiferromagnetic spin chains in an optical lattice},\ }\href
  {https://doi.org/10.1038/nature09994} {\bibfield  {journal} {\bibinfo
  {journal} {Nature}\ }\textbf {\bibinfo {volume} {472}},\ \bibinfo {pages}
  {307–312} (\bibinfo {year} {2011})}\BibitemShut {NoStop}%
\bibitem [{\citenamefont {\change{Shiraishi, Naoto and Mori,
  Takashi}}(2017)}]{Mori2017}%
  \BibitemOpen
  \bibfield  {author} {\bibinfo {author} {\bibnamefont {\change{Shiraishi,
  Naoto and Mori, Takashi}}},\ }\bibfield  {title} {\bibinfo {title}
  {\change{Systematic Construction of Counterexamples to the Eigenstate
  Thermalization Hypothesis}},\ }\href
  {https://doi.org/10.1103/PhysRevLett.119.030601} {\bibfield  {journal}
  {\bibinfo  {journal} {Phys. Rev. Lett.}\ }\textbf {\bibinfo {volume} {119}},\
  \bibinfo {pages} {030601} (\bibinfo {year} {2017})}\BibitemShut {NoStop}%
\bibitem [{\citenamefont {Fishman}\ \emph {et~al.}(2022)\citenamefont
  {Fishman}, \citenamefont {White},\ and\ \citenamefont
  {Stoudenmire}}]{ITensor}%
  \BibitemOpen
  \bibfield  {author} {\bibinfo {author} {\bibfnamefont {M.}~\bibnamefont
  {Fishman}}, \bibinfo {author} {\bibfnamefont {S.~R.}\ \bibnamefont {White}},\
  and\ \bibinfo {author} {\bibfnamefont {E.~M.}\ \bibnamefont {Stoudenmire}},\
  }\bibfield  {title} {\bibinfo {title} {The {ITensor} software library for
  tensor network calculations},\ }\href
  {https://doi.org/10.21468/SciPostPhysCodeb.4} {\bibfield  {journal} {\bibinfo
   {journal} {SciPost Phys. Codebases}\ ,\ \bibinfo {pages} {4}} (\bibinfo
  {year} {2022})}\BibitemShut {NoStop}%
\bibitem [{\citenamefont {Schecter}\ and\ \citenamefont
  {Iadecola}(2018)}]{Iadecola2018a}%
  \BibitemOpen
  \bibfield  {author} {\bibinfo {author} {\bibfnamefont {M.}~\bibnamefont
  {Schecter}}\ and\ \bibinfo {author} {\bibfnamefont {T.}~\bibnamefont
  {Iadecola}},\ }\bibfield  {title} {\bibinfo {title} {Many-body spectral
  reflection symmetry and protected infinite-temperature degeneracy},\ }\href
  {https://doi.org/10.1103/PhysRevB.98.035139} {\bibfield  {journal} {\bibinfo
  {journal} {Phys. Rev. B}\ }\textbf {\bibinfo {volume} {98}},\ \bibinfo
  {pages} {035139} (\bibinfo {year} {2018})}\BibitemShut {NoStop}%
\bibitem [{\citenamefont {Bull}\ \emph {et~al.}(2019)\citenamefont {Bull},
  \citenamefont {Martin},\ and\ \citenamefont {Papi\ifmmode~\acute{c}\else
  \'{c}\fi{}}}]{Papic2019}%
  \BibitemOpen
  \bibfield  {author} {\bibinfo {author} {\bibfnamefont {K.}~\bibnamefont
  {Bull}}, \bibinfo {author} {\bibfnamefont {I.}~\bibnamefont {Martin}},\ and\
  \bibinfo {author} {\bibfnamefont {Z.}~\bibnamefont
  {Papi\ifmmode~\acute{c}\else \'{c}\fi{}}},\ }\bibfield  {title} {\bibinfo
  {title} {Systematic construction of scarred many-body dynamics in 1d lattice
  models},\ }\href {https://doi.org/10.1103/PhysRevLett.123.030601} {\bibfield
  {journal} {\bibinfo  {journal} {Phys. Rev. Lett.}\ }\textbf {\bibinfo
  {volume} {123}},\ \bibinfo {pages} {030601} (\bibinfo {year}
  {2019})}\BibitemShut {NoStop}%
\bibitem [{\citenamefont {D'Alessio}\ \emph {et~al.}(2016)\citenamefont
  {D'Alessio}, \citenamefont {Kafri}, \citenamefont {Polkovnikov},\ and\
  \citenamefont {Rigol}}]{D'Alessio2016}%
  \BibitemOpen
  \bibfield  {author} {\bibinfo {author} {\bibfnamefont {L.}~\bibnamefont
  {D'Alessio}}, \bibinfo {author} {\bibfnamefont {Y.}~\bibnamefont {Kafri}},
  \bibinfo {author} {\bibfnamefont {A.}~\bibnamefont {Polkovnikov}},\ and\
  \bibinfo {author} {\bibfnamefont {M.}~\bibnamefont {Rigol}},\ }\bibfield
  {title} {\bibinfo {title} {From quantum chaos and eigenstate thermalization
  to statistical mechanics and thermodynamics},\ }\href
  {https://doi.org/10.1080/00018732.2016.1198134} {\bibfield  {journal}
  {\bibinfo  {journal} {Adv. Phys.}\ }\textbf {\bibinfo {volume} {65}},\
  \bibinfo {pages} {239} (\bibinfo {year} {2016})}\BibitemShut {NoStop}%
\bibitem [{\citenamefont {Oganesyan}\ and\ \citenamefont
  {Huse}(2007)}]{Oganesyan2007}%
  \BibitemOpen
  \bibfield  {author} {\bibinfo {author} {\bibfnamefont {V.}~\bibnamefont
  {Oganesyan}}\ and\ \bibinfo {author} {\bibfnamefont {D.~A.}\ \bibnamefont
  {Huse}},\ }\bibfield  {title} {\bibinfo {title} {Localization of interacting
  fermions at high temperature},\ }\href
  {https://doi.org/10.1103/PhysRevB.75.155111} {\bibfield  {journal} {\bibinfo
  {journal} {Phys. Rev. B}\ }\textbf {\bibinfo {volume} {75}},\ \bibinfo
  {pages} {155111} (\bibinfo {year} {2007})}\BibitemShut {NoStop}%
\bibitem [{\citenamefont {Santos}\ and\ \citenamefont
  {Rigol}(2010)}]{Rigol2010}%
  \BibitemOpen
  \bibfield  {author} {\bibinfo {author} {\bibfnamefont {L.~F.}\ \bibnamefont
  {Santos}}\ and\ \bibinfo {author} {\bibfnamefont {M.}~\bibnamefont {Rigol}},\
  }\bibfield  {title} {\bibinfo {title} {Onset of quantum chaos in
  one-dimensional bosonic and fermionic systems and its relation to
  thermalization},\ }\href {https://doi.org/10.1103/PhysRevE.81.036206}
  {\bibfield  {journal} {\bibinfo  {journal} {Phys. Rev. E}\ }\textbf {\bibinfo
  {volume} {81}},\ \bibinfo {pages} {036206} (\bibinfo {year}
  {2010})}\BibitemShut {NoStop}%
\bibitem [{\citenamefont {Karle}\ \emph {et~al.}(2021)\citenamefont {Karle},
  \citenamefont {Serbyn},\ and\ \citenamefont {Michailidis}}]{Karle}%
  \BibitemOpen
  \bibfield  {author} {\bibinfo {author} {\bibfnamefont {V.}~\bibnamefont
  {Karle}}, \bibinfo {author} {\bibfnamefont {M.}~\bibnamefont {Serbyn}},\ and\
  \bibinfo {author} {\bibfnamefont {A.~A.}\ \bibnamefont {Michailidis}},\
  }\bibfield  {title} {\bibinfo {title} {Area-law entangled eigenstates from
  nullspaces of local {Hamiltonians}},\ }\href
  {https://doi.org/10.1103/PhysRevLett.127.060602} {\bibfield  {journal}
  {\bibinfo  {journal} {Phys. Rev. Lett.}\ }\textbf {\bibinfo {volume} {127}},\
  \bibinfo {pages} {060602} (\bibinfo {year} {2021})}\BibitemShut {NoStop}%
\end{thebibliography}
\end{document}